\documentstyle[emulateapj,epsf]{article} 

\begin{document}
\submitted{Accepted by The Astrophysical Journal} 
\title
{A gas-rich nuclear bar fuelling a powerful central starburst in NGC~2782}
\author {Shardha Jogee\altaffilmark{1}, Jeffrey D. P. Kenney\altaffilmark{2}
\& Beverly J. Smith\altaffilmark{3}}
\authoremail{sj@phobos.caltech.edu, kenney@astro.yale.edu, beverly@ipac.caltech.edu}
\altaffiltext{1}{Division of Physics, Mathematics, and Astronomy, MS 105-24, 
California Institute of Technology, Pasadena, CA 91125}
\altaffiltext{2}{Yale University Astronomy Department, New Haven, CT 06520-8101}
\altaffiltext{3}{ IPAC/Caltech, Pasedena, CA 91125}

\begin{abstract}
We present evidence that the peculiar interacting 
starburst galaxy NGC~2782 (Arp 215) harbors a gas-rich nuclear stellar bar 
feeding an M82-class powerful central starburst, from a study based on 
high resolution interferometric  CO (J=1-$>$0) data, optical BVR and 
H$\alpha$ observations, along 
with available NIR images, a 5 GHz radio continuum map, and 
archival HST images.
Morphological and kinematic data show that 
NGC~2782 harbors a  clumpy, bar-like  CO  feature 
of radius $\sim$ $7.5''$ 
(1.3 kpc) which  leads  a nuclear stellar bar of similar size.   
The nuclear   bar-like CO  feature is massive: 
it contains  $\sim$ $2.5 \times 10^9$ M$_{\tiny \odot}$ of molecular 
gas,   which makes  up $\sim$~8~\%  of the dynamical mass 
present within  a  1.3 kpc radius. Within the CO bar, emission 
peaks in two extended clumpy lobes which lie on opposite sides 
of the nucleus, separated by $\sim$ $6''$ (1 kpc). 
Between the CO lobes, in the inner 200 pc radius, resides 
a powerful  central starburst which is forming stars 
at a rate of 3 to 6  M$_{\tiny \odot}$ yr$^{-1}$. 
While circular motions dominate the  CO velocity field, 
the CO lobes show weak bar-like streaming motions on 
the leading side of the nuclear stellar bar, suggestive of gas
inflow. We estimate  semi-analytically   the  gravitational torque 
from the nuclear  stellar bar on the gas, and suggest
large gas inflow rates from the CO lobes into the central starburst. 
These  observations,  which   are  amongst  the  first ones 
showing  a  nuclear stellar bar  
fuelling molecular  gas into an intense central starburst,  
are  consistent with simulations 
and theory which  suggest that nuclear bars provide  an efficient way  
of  transporting gas closer to the galactic center to fuel central 
activity.  Furthermore, several massive clumps ($10^7$ - $10^8$ M$_{\tiny \odot}$) 
are present at low radii, and dynamical friction might 
produce further gas inflow. 
We suggest that the nuclear bar-like  molecular gas feature 
and central activity will be very short-lived, 
likely disappearing  within $5 \times 10^8$ years.

\end{abstract}


\section{Introduction}
In order to drive gas from the outer disk 
to the nucleus of a spiral galaxy,  from scales of tens of kpc 
to tens of pc, the angular momentum of the gas must be reduced by 
around six orders of magnitude. 
It is now accepted that a wide variety of phenomena involving 
torques and dissipation  can drive gas from the outer disk 
into the circumnuclear (inner kpc) region. 
Non-axisymmetric features (Simkin et al. 1980) 
such as  isolated and induced primary bars   
(Schwarz 1984; Noguchi 1988), as well as 
tidal interactions and mergers (Negroponte \& White 1983; 
Mihos \& Hernquist 1994) can be relevant. 
However, it is less clear how gas can be driven from scales 
of kpc to tens of pc. 
Shocks associated with the primary bar tend to weaken in the 
inner part of the galaxy and cannot drive further inflow 
(Shlosman et al. 1989). 
Numerical simulations (e.g., Combes \& Gerin 1985) 
and observations (e.g., Kenney et al. 1992) 
show that gas often piles up in rings near  inner Lindblad 
resonances  (ILRs). 
Some studies  (e.g., Combes \& Gerin 1985) 
have suggested that gravity torques reverse near ILRs, and thereby  stall the transport of gas 
towards the galactic center. 
In order to overcome this  transport barrier, 
theory  (Shlosman et al. 1989) 
and simulations (Friedli \& Martinet 1993; Combes 1994; Heller \& 
Shlosman 1994) have suggested  that nuclear bars 
nested within  a primary bar (or  `bars within 
bars')  can be  efficient mechanisms for 
driving gas closer to the galactic center to  
fuel  central  starbursts and AGNs.

From the observational standpoint, 
there is mounting evidence from optical and NIR surveys 
that a large number of galaxies have nuclear stellar 
bars  (Buta \& Crocker 1993; Wozniak et al. 1995; Shaw et al. 1995;  Elmegreen et al.  1996; Friedli et al. 1996; 
Jungwiert et al. 1997), and  a significant fraction of these galaxies 
show nuclear  activity. 
However, to date,  severely lacking  are observations
which map the molecular gas in these systems, and show the 
nuclear stellar bar feeding the  gas into a central 
starburst or AGN. Such observations are highly desirable 
to constrain formation scenarios and typical lifetimes of 
nuclear bars, as well as to  assess their relative importance  with 
respect to other fuelling mechanisms.
The current work  presents evidence 
that the peculiar  galaxy NGC~2782 (Arp 215)
harbors a massive bar-like molecular feature of 
radius  1.3 kpc and  a  similarly-sized nuclear stellar bar 
which appears to be fuelling  molecular gas 
into  the inner 200 pc radius, where resides a powerful 
central starburst, comparable  in intensity to the archetypal starburst 
M82. 
This  study is based on  high resolution ($2''$) interferometric
Owens Valley Radio Observatory (OVRO) CO (J=1-$>$0) data, 
optical observations from 
the Wisconsin Indiana Yale NOAO (WIYN) telescope, along with 
NIR images from Engelbracht, Rieke, \& Rieke (1999),  
a 5 GHz radio continuum (RC) map  from Saikia et al. (1994), and archival 
optical HST images.  This paper is  organised as follows. 
$\S$~2 
\rm
outlines the previous  work done on NGC 2782 and 
$\S$~3 
\rm
describes  the new observations. 
$\S$~4.1 
\rm
presents evidence for a  nuclear stellar bar and a possible 
large-scale primary stellar bar. 
$\S$~4.2-4.4 
\rm
present evidence for  a nuclear bar-like CO feature 
of similar size as the nuclear stellar bar, and  
discuss its kinematics, inclination, and  mass fraction.
$\S$~4.5
\rm
describes  the central starburst and how it 
relates  to the nuclear bars.
$\S$~4.6 
\rm
compares  the  CO bar in    NGC 2782 to other
systems which host  circumnuclear bar-like CO 
morphologies.
$\S$~4.7 
\rm
discusses the evolution of  the circumnuclear region. 
At what rate is gas being fuelled into the  
central starburst? 
How long will the central activity and nuclear bars  last?
In  
$\S$~4.8, 
\rm
we compare NGC 2782 to theory and simulations in order 
to investigate  how the  nuclear stellar and gas bars  
formed.

\section {Properties of  NGC~2782 from previous studies}

NGC~2782  (Arp 215), classified as  
peculiar SABa(rs) (de Vaucouleurs et al. 1991, in RC3)
or Sa(s) Peculiar (Sandage \& Tammann 1981), 
is highly suited for addressing numerous astrophysical questions: 
\bf
(a)\rm~It harbors  one of the most luminous 
circumnuclear starbursts among nearby (D $<$40 Mpc) spirals 
(Devereux 1989). The starburst  is comparable to the one in M82, 
with a FIR luminosity  of $2 \times 10^{10}$ L$_{\sun}$ 
(Smith 1991), and a central 10 micron luminosity 
of $1 \times 10^9$ L$_{\sun}$ (Giuricin et al. 1994). 
\bf
(b)\rm~NGC 2782 is both moderately inclined and  
nearby (D = 34 Mpc for H$_{\rm o}$ of 75 km s$^{-1}$ Mpc $^{-1}$), 
and therefore the molecular gas can 
be  resolved on scales of a few  hundred parsecs. 
This is an important advantage over many  distant or highly inclined 
luminous interacting galaxies 
(e.g.,  Bryant   \& Scoville 1999; Evans et al. 1999;  Yun et al. 1995) 
which have been previously mapped in CO.
\bf
(c)
\rm
The large scale  properties (see below) 
of NGC~2782  suggest it is the product of 
an intermediate mass-ratio interaction or merger. 
While minor and major interactions have been widely 
targeted by both simulations 
(e.g., Hernquist \& Mihos 1995; Barnes \& Hernquist 1996)
and  observations, studies of  intermediate mass-ratio 
interactions/mergers remain comparatively rare.

\it
Large-scale HI and optical properties of NGC 2782:
\rm
NGC~2782 has a pair of prominent HI and optical tails 
(Smith 1991; Sandage \& Bedke 1994) which bracket an
optical disk that harbors three ripples. The eastern
tail has been detected in CO (Smith et al. 1999). 
The R-band image in Fig. 1 (Plate 1) shows the optical disk and tails. 
Except for the three
ripples at radii of $\sim$ $25''$, $45''$, and $60''$, 
the optical disk is relatively undisturbed within a 
radius of 1 $'$  (10 kpc). 
The optical disk can be fitted with an exponential surface 
brightness profile between  radii of $\sim$ $30''$ to $60''$ 
(e.g., Smith 1994). 
This shows the galaxy is not as heavily disturbed as major mergers 
where violent relaxation leads to an  r$^{1/4}$ surface brightness 
profile rather than an exponential one.
On the other hand, the presence of the 
two prominent large-scale HI tails suggests that NGC~2782 
is not a minor interaction/merger either. 
In fact, the  optical and HI properties have been modeled by Smith (1994) 
with an interaction of two disk galaxies 
which have an intermediate (1:4) mass ratio, and reached closest approach 
$\sim$ $ 2 \times 10^8$ years ago. An alternative possibility is that 
NGC~2782 is a merger of two disks of unequal mass. 

\it
Circumnuclear properties of NGC~ 2782:
\rm
Low resolution H$\alpha$ observations (Hodge \& Kennicutt 
1983; Smith 1994; Evans et al. 1996) show an unresolved 
bright region of star formation in the inner $8''$ 
(1.4 kpc) radius. This region has  optical emission spectra 
(Sakka et al. 1973; Balzano 1983; Kinney et al. 1984) 
indicative of HII regions,  as well as an 
additional component of  highly excited, highly 
ionised gas  which is part of a nuclear outflow
(Kennicutt et al. 1989;  Boer et al. 1992). 
Jogee, Kenney, \& Smith (1998) hereafter 
\bf 
Paper I, 
\rm 
used optical images, 5 GHz RC data  and X-ray data 
to resolved the inner few kpc  into 
(i) a centrally peaked region of star formation which is 
elongated east-west with dimensions $\sim$  $ 1 \times 1.7$ 
kpc, 
(ii) a starburst-driven outflow which forms 
a well-defined  collimated  bubble having an extent of 
$\sim$ 1 kpc, and a closed shell at its edge, as seen in H$\alpha$, 
[O~III], and 5 GHz RC. 
While starburst-driven outflows are common, such 
a remarkable bubble morphology is relatively rare 
in other starburst galaxies of comparable luminosity. 
We argued, based on the outflow morphology and timescale 
of $\sim$  $4  \times$  10$^{6}$ years (Boer et al.  1992), that this outflow 
is dynamically younger than biconical, freely-expanding outflows 
seen in other starburst galaxies of comparable  IR luminosity 
(e.g., M82). We presented evidence that the outflow is driving 
hot and warm  ionised gas, and possibly cold molecular gas, out 
of the central kpc of the galaxy.

\section{Observations}

\subsection{CO observations}
The central $65''$ of NGC 2782 was observed in the 
CO (J=1-$>$0) transition at 115.27 GHz with the OVRO millimeter-wave 
interferometer (Padin et al. 1991) between February and May 1995.
The array consists of six 10.4 m telescopes with primary half power 
beam width of $65''$ at 115 GHz . 
The galaxy was observed in two array configurations 
with projected baselines ranging from 15 to 242 m. 
Data were obtained simultaneously 
with an analog continuum correlator of bandwidth 1 GHz and 
a digital spectrometer which is made of four independent modules that 
each have 
32 channels with a velocity resolution of 10.4 km s$^{-1}$. For our
observations, the modules were partially overlapping and covered a 
total bandwidth of 460 MHz with 116 frequency channels. The passband
and flux calibration of the data were carried out using the 
Owens Valley millimeter array software  (Scoville et al. 1993). We 
corrected for temporal phase variation by observing 
the phase calibrator 0923+392 every 30 minutes.
We corrected for gain variation from one channel to the next 
across the spectrometer band using spectra of the bright quasars
3C273, 3C84, and 3C345. The absolute flux scale was determined 
from observations of  Uranus, Neptune, and 3C273.
We used the NRAO AIPS software to Fourier transform the calibrated
uv data and deconvolve the channel maps with the `CLEAN' algorithm
as implemented in the AIPS task `MX'. We made both  
uniformly weighted and naturally weighted channel maps. For the 
uniformly weighted maps,  the size of 
the synthesized beam was $2.1'' \times 1.5''$ ($355 \times 255 $ pc ), 
at a position angle (PA) 
of -89.8 $\deg$, the typical r.m.s. noise was 13 mJy per beam, 
and the peak signal-to-noise was 7. 
For the naturally weighted maps, the corresponding 
quantities were  $2.6'' \times 2.0''$ ($440 \times 340$  pc) at a 
PA of 89.7 $\deg$, 10.5 mJy per beam, and 12.5 respectively.
The naturally weighted channel maps are shown in Fig. 2. 
We detect emission above the 3 $\sigma$ level in 35 
channels which span velocities ranging from 2365 to 2730 km s$^{-1}$. 
The naturally weighted maps detect a total flux of 195 
Jy km s$^{-1}$  from the central $40''$ (7 kpc) diameter. 
This corresponds to 
73~\% of the single dish FCRAO flux within $45''$ (Young et al. 
1995), and 1.7 times the flux detected in the uniformly 
weighted channel maps. 
We combined the clean channels showing emission to make moment 0, 1, 
and 2 maps, which represent the total intensity, 
the intensity-weighted velocity field,  and the 
velocity dispersion field respectively.

\subsection{Optical observations}
We refer the reader to Paper I  for 
the observations and reduction of the optical BVR and  H$\alpha$ 
images taken on the WIYN telescope. 
For the archival HST images, the original  
observations were made on April 18, 1997 using the Wide 
Field Planetary Camera 2 (WFPC2) with a scale  of  0.0455"/pix.  
Two  exposures of 230 s  were taken in 
each  of the broadband  F555W and F814W filters, 
which correspond approximately to the V and  I bands 
respectively. One 600 s exposure
and two 1300 s exposures  were also taken 
in the  narrowband FR680N filter.
This  filter is a tunable linear ramp filter,
whose central wavelength varies with position across the
chip.  For these observations, the nucleus of NGC 2782 was placed in the 
PC chip and the filter
was rotated +15$^{\circ}$ from its nominal position
(the FR680P15 mode).  With this orientation,
the central core of NGC 2782 was observed with
an effective 
central wavelength of 6623$\AA$ and a bandwidth of 86$\AA$.  
This bandpass
contains the redshifted H$\alpha$ line and the 6584$\AA$ [N~II] line of
NGC 2782.  With this filter, the usable field of view
is  $\sim$ $10''$  because of vignetting and cross-talk effects.
The images were cleaned of cosmic rays and multiple images in each filter 
were summed.

\section{Results and Discussion} 

\subsection{The nuclear stellar bar and large-scale oval}
Figure 3a shows the I-band image (Smith 1994) 
with a 2$'$ (20 kpc) field of view. The very outermost isophotes 
are relatively circular, and surround a weak oval feature 
which has  a position angle (PA) of $\sim$~20 $\deg$ and 
 a radius $\sim$ $25''$ (4.3 kpc). 
This oval  feature is flanked by two 
relatively straight dust lanes which are visible in the B and B-V images 
(Figs. 7a \& 7b, Plate 2). 
The locations  of the dust lanes  are shown as  solid
lines in Fig. 3a. 
The  dust lanes  appear to extend between  radii of $8''$  to $25''$, 
and are offset towards the leading edge of the large-scale 
oval feature. The appearance of the dust lanes is similar to 
the dust lane morphologies observed along the leading edges of 
primary stellar bars in many spiral galaxies e.g., 
NGC~1300 (Sandage 1961), NGC~1365 (Teuben et al. 1986). 
Figure 3b shows the inner 1$'$ (10 kpc)  diameter of the K-band 
image from Engelbracht et al. (1999).
The uniformly-weighted CO map is superposed as greyscale, and 
solid lines again mark the locations  of the dust lanes. 
Between the dust lanes, in the inner $7.5''$ (1.3 kpc) 
radius, the K-band image 
shows a nuclear oval feature at  a  PA of 
$\sim$ 100 $\deg$.

Is the large-scale oval feature of radius $\sim$ 4.3 kpc a primary bar, 
and the inner oval of radius $\sim$ 1.3 kpc 
a nuclear stellar bar? 
We perform an isophotal analysis of the K-band and I-band images 
to  address this question. 
The results are shown in Fig. 4, where  diamonds and 
crosses represent  the K-band and I-band data respectively: 
\bf 
(a) 
\rm 
The ellipticity profile of the K-band light 
shows an inner ellipticity maximum (0.28) over
which the PA maintains an approximately constant 
value (100 $\deg$). Beyond a radius of 
$\sim$ $7.5''$ (1.3 kpc), the ellipticity 
falls to a minimum  (e $<$ 0.1 ) 
and the position angle changes abruptly from 100  $\deg$ to 
0-10  $\deg$. 
\bf
(b) 
\rm
Between radii of $8''$  to $\sim$~$25''$ (4.3 kpc), 
the ellipticity  smoothly rises 
to  0.26  while the position  angle twists gradually 
to a PA of $\sim$ 20 $\deg$. 
Further out,  given the low  signal-to noise 
in the K-band image, we use the I-band image, where 
the ellipticity  falls from 0.26  to  below  0.1 
in the outer disk, at radii$>$ $50''$ (8.5 kpc).

The  relatively constant position angle (100 $\deg$) 
of the nuclear oval and its  inner ellipticity maximum (0.28), 
which is significantly higher than the outer disk's ellipticity 
($<$ 0.1), strongly suggest the existence of 
a nuclear stellar bar  of radius $7.5''$ (1.3 kpc).  
Previous NIR  studies  (Pompea \& Rieke 1990; Forbes et al. 1992) 
hinted at the existence of this nuclear stellar bar, but it 
was not  as  evident because of lower sensitivity and resolution. 
As further supporting evidence, we 
note that for a moderately inclined galaxy like NGC~2782, 
the nuclear oval is unlikely to be caused   by 
projection artifacts which may  sometimes emulate a bar 
 (e.g., Friedli et al. 1996; Jungwiert et al. 1997). 
In addition, the  maximum  ellipticity (0.28)  of  the nuclear
stellar oval is comparable  to the mean maximum ellipticity (0.31) found 
for the inner cores of other galaxies considered to have nuclear stellar bars 
(e.g., Friedli et al. 1996).
The presence of a nuclear stellar bar is also consistent with 
the difference between the  PA of the nuclear stellar oval, 
and the PA of the line of nodes, as  estimated 
from the properties of  the starburst-driven outflow. 
The RC bubbles and the H$\alpha$ bubble 
associated with the outflow (Paper I, also Fig. 3d) suggest 
a kinematic minor axis of 165 $\deg$, and  
for an outflow  perpendicular to the disk, 
this sets the line of nodes  at $\sim$ 75 $\deg$. 
The HST I and V-band images (see $\S$ 4.5), 
more obscured  by dust than the K-band image, 
also reveal stellar and dust lane morphologies  consistent with 
the existence of a nuclear stellar bar. 

While the case for a nuclear stellar bar of radius $7.5''$ (1.3 kpc) 
seems firm, we cannot unambiguously determine if the large-scale oval 
feature which shows a smooth rise in ellipticity from a radius of  
$8''$  to   $25''$ (4.3 kpc)    is a primary bar.
On one hand, 
the relatively straight and prominent dust lanes (described above) 
associated with the leading edge of the  large-scale oval  are strongly 
reminiscent of primary stellar bars.  
On the other hand, we note that  within the large-scale oval, 
between radii of  $8''$  to $25''$, 
the PA  of the I-band isophotes is not constant, but twists. 
A constant position angle would have clinched the  case 
for a primary bar, but  isophotal twist neither proves nor 
excludes the presence of a primary bar.
In general, spiral arms,  dust lanes, 
a radius-dependent triaxiality associated with a primary bar 
or/and a bulge,  as well as  intrinsically misaligned isophotal 
surfaces not associated with a primary bar,  may all produce 
isophotal twist (Wozniak et al. 1995; Jungwiert et al. 1997). 
In the case of NGC 2782, we can exclude  spiral arms 
or ripples as the cause of the observed  isophotal twist, since 
such features are  not present between radii  of $8''$  to $25''$  
in the K-band or I-band image (Figs. 3a-b). 
(The ripples in the I-band image exist only further out, 
at radii of  $\sim$ $25''$, $45''$, and $60''$).
However, since dust and other causes of isophotal twist cannot be excluded, 
a firm case cannot be made for a primary stellar bar. 
In summary, we hence conclude that NGC 2782 
harbors a nuclear stellar bar  of  radius $7.5''$ (1.3 kpc), 
which is nested within a large-scale oval of radius  $25''$ (4.3 kpc),  
which  may be, but is not necessarily, a primary bar.

\subsection{The circumnuclear CO morphology and kinematics}


The  previously published low resolution ($6''$ or 1 kpc) CO (J=1-$>$0) 
Nobeyama Millimeter Array  (NMA) map  (Ishizuki 1994) 
shows an elongated feature  with two barely  resolved CO peaks 
in the inner  $9''$ (1.5 kpc) radius, and  fainter emission further 
out. The data gave  limited additional information  on the distribution 
and kinematics of the gas in the inner few kpc.
Our  high resolution ($2.1'' \times 1.5''$ or $355 \times 255 $ pc)  OVRO 
CO (J=1-$>$0)  map in Fig. 5a  shows a    bar-like  CO feature 
of radius $\sim$~$7.5''$ (1.3 kpc), and resolves it into 
two extended, clumpy lobes which 
lie on opposite sides of the nucleus, separated by $\sim$ $6''$ (1 kpc). 
Between the two CO lobes, in the inner 200 pc radius, the 
starburst activity peaks sharply  (Paper I  \& $\S$ 4.5).
It is possible that the CO depression between the lobes is the result 
of  gas consumption by star formation or/and gas blown out 
in the starburst-driven wind.
The  bar-like CO feature has a similar size as the nuclear stellar 
bar of radius $7.5''$ (1.3 kpc)  and PA $\sim$ 100 $\deg$ identified 
in  $\S$ 4.1. 
The two CO lobes lie on the leading edges of this nuclear stellar bar, 
assuming a clockwise sense of rotation for the stellar bar. 
(This  sense of rotation for the stars stems from the assumption that 
they  are rotating in the same sense as the molecular gas, 
for which we infer   a clockwise sense of rotation  from the kinematics). 
Gas on the leading edges of  a stellar bar is 
expected to lose angular momentum and be driven inwards, 
as a result of shocks and  gravitational torques exerted by 
the stars on the gas (e.g.,  Schwarz  1984; 
Combes \& Gerin 1985;  Athanassoula 1992; 
Friedli \& Martinet 1993; Byrd et al 1994; Piner et al. 1995).  
It is thus possible that the nuclear stellar bar 
is fuelling gas from the CO lobes into the central  
starburst.

The intensity-weighted CO velocity field shown in Fig. 5b 
provides further insight into the circumnuclear dynamics. 
No  significant sign of warping is seen : there is no 
conspicuous  distortion of 
the kinematic  principal axes into S-shapes over the inner 
$8''$  (1.4 kpc) radius , and  the PA of the   CO kinematic 
minor axis  (KMNA)  is consistent with PA  (165 $\deg$) 
of the minor axis suggested by the starburst-driven 
RC  and  H$\alpha$ outflow bubbles 
(Paper I \&  Fig. 3d). 
Most of the  isovelocity contours in the  central $8''$ 
(1.4 kpc) radius  form a  `spider-diagram', 
suggesting that circular motions are important. 
The CO velocity field, with redshifted velocities lying to the 
west of the kinematic center, implies a clockwise sense of 
rotation for the circumnuclear molecular gas 
if we assume the near side of the galactic disk lies to 
the north. This orientation for the galactic disk was inferred 
in Paper I from the morphology of the starburst-driven outflow and the 
dust lanes (see also  $\S$ 4.3).
The circumnuclear CO  velocities are consistent with the previously 
observed kinematics of the cores of  HI (Smith 1994) and 
ionised (Boer et al. 1992) gas.
However, superposed on the predominantly circular velocity field, 
are non-circular  motions in several 
regions, particularly near the CO lobes. 
The kinematic major axis  (KMJA) is at $\sim$ 
75 $\deg$ near the center, but it curves near the eastern 
and western CO lobes.  This suggests bar-like streaming motions in the 
CO lobes.

The CO spatial velocity plot along the KMJA, 
shown in Fig. 6a,  probes the gas kinematics in more detail. 
The spatial velocity plot has a width of $\pm$ $1''$ about the 
KMNA, and the systemic velocity of 2550 km s$^{-1}$ has been 
subtracted from the velocities shown on the y-axis.
On the eastern side of the  major axis, between 
$1''$ and  $4''$,  complex kinematics 
and an apparent gap in the 
spatial-velocity  plot at a velocity of -50  km s$^{-1}$ 
are  seen .   
The features marked as A1, A2, A3, and A4  in Fig. 6a 
delineate  large linewidths at the position of 
the western and main  eastern CO peaks.   
Notice also that the observed velocities increase 
rapidly between the center and A3, but  the  `rotation curve' 
become shallow as it goes through the western  CO peak.
What is producing the complex kinematics in the CO peaks? 
These kinematics may be the manifestation of 
bar-like azimuthal streaming  motions  and  radial inflow motions 
which are expected in gas lying  on the leading edges 
of a nuclear stellar bar. 
It is difficult to  directly  identify a radial inflow component 
in the data  because  the observed line of sight velocity 
is in general a combination of several velocity components 
which cannot be easily disentangled from each other: 
azimuthal in-plane motions, vertical out-of-plane motions, and   
radial  in-plane  motions (Eq. 8-60  in Mihalas \& Binney 1981). 
In the case of NGC 2782, spatial velocity  cuts 
which go through the CO peaks are close to the KMJA, 
and hence are  more sensitive to azimuthal in-plane 
and vertical out-of-plane motions than to radial  in-plane  motions.

We also point out 
that some molecular gas might be outflowing from the starburst 
region. The CO intensity map in Fig. 5a shows 
two CO spurs (labelled O1 and O2) between the  CO lobes.
The CO isovelocity contours in Fig. 5b 
show  clear kinks at the base of these spurs, indicating 
deviations from circular motions. 
Fig. 3d shows that these spurs lie just north and south of  
the central starburst region, and are elongated along 
the CO KMNA ($165\deg$), within the   two RC outflow bubbles. 
In Fig. 6b, the spatial-velocity plot along the KMNA 
shows that gas velocities deviate from the 
systemic velocity (2555 km s$^{-1}$) by 
$\sim$ (+ 30 km s$^{-1}$) in the northern spur, and by 
$\sim$ (- 30 km s$^{-1}$) in the southern spur. 
With the near side of the disk being north 
(Paper I \& $\S$ 4.5), these velocities are consistent with 
vertical out-of-plane motions or radial  in-plane  motions. 
Although we cannot strictly distinguish between the 
two possibilities, the location of the spurs 
with respect to the central starburst and outflow bubbles, 
and their elongation along the KMNA  
provide circumstantial support for outflowing gas.

\subsection{The inclination of the circumnuclear molecular gas}

We need  to estimate the inclination  of the circumnuclear molecular gas  
in order to  subsequently constrain  properties such as 
the molecular gas mass fraction and  kinematics.
We first note that the CO velocity field (Fig. 5b) does not suggest 
any significant warping,  and therefore it is reasonable to assume 
that the circumnuclear molecular gas  lies in one plane. 
We infer that  the  inclination  (i$_{\rm co}$) of this plane 
is $\sim$ 30 $\deg$, from the following arguments: 
\bf
(i)
\rm 
We can rule out an  inclination close to edge-on  because 
it  would imply that the large thickness of the observed 
bar-like CO distribution corresponds to  unrealistically large 
scale heights ($\sim$ 1 kpc) for  molecular gas    \it 
throughout the region \rm 
 between radii of $3''$ (510 pc) to $7.5''$ (1.3 kpc).
While one might conceivably justify large scale heights in the central 
$2''$ (340 pc) radius where intense starburst activity 
and winds prevail, it would be difficult to account 
for such large scale heights out to a radius of  1.3 kpc. 
\bf
(ii)
\rm 
The Tully Fisher relation for optical blue magnitudes requires 
i$_{\rm co}$  $>$   25 $\deg$ in order to ensure that the 
peak CO rotational speed (165 km $^{-1}$/sin $i_{\rm co}$) 
does not exceed the maximum rotational speed   of 400 km s$^{-1}$ 
(Rubin et al. 1985) observed  in peculiar and normal galaxies
which are  optically as bright or brighter than NGC~2782. 
We can also apply  the infrared Tully Fisher  relation between 
the   H magnitude (m$_{\rm H}$) and  the peak  rotational speed  
($V_{\rm max}$), [m$_{\rm H}$ = 2 - log V$_{\rm max}$] 
(Pierce \& Tully 1988). 
Assuming $m_{\rm H}$ =  10.1 (de Vaucouleurs \& Longo  1988), 
and an  uncertainty of  0.7 magnitude in  the Tully Fisher relation, 
we find that  $i_{\rm co}$   is  ($31 \pm 6 \deg$). 
\bf
(iii)
\rm
This moderate value of  $\sim$ 30 $\deg$  we  estimated 
for $i_{\rm co}$  is 
consistent with the inclination of  dust in the inner few kpc, 
as constrained by properties  of the starburst-driven outflow. 
The 5 GHz RC image shows 
kpc-sized outflow bubbles \it both \rm  north and south of the nucleus, 
while only the   southern bubble is visible 
in the optical H$\alpha$ and [O~III] images (see Paper I). 
This suggests that the northern bubble is obscured  
by dust  in the inner few kpc of 
a moderately inclined disk whose  near side  lies 
to the north (Jogee et al. 1998). 
\bf
(iv)
\rm
Based on the relatively undisturbed  appearance 
of the outer disk (Fig. 1; Plate 1) 
within a radius of $1'$ (10 kpc), one  expects 
its inclination  to be similar to i$_{\rm co}$. 
It is reassuring to note that 
the  outermost isophotes of the I-band image  (Fig. 3a \& 4) 
do indeed suggest an inclination of $\sim$ 30 $\deg$ for the outer 
disk.

How do the nuclear stellar bar and the 
bar-like CO feature in the inner $7.5''$ (1.3 kpc) radius  
relate to the large-scale features  in NGC 2782? 
The naturally weighted OVRO  CO map 
is overlaid on the B-V image  in Fig 7b. (Plate 2). 
The bar-like CO feature   of radius $7.5''$ (1.3 kpc)  
contains  75~\% of the total CO emission detected 
within the  central $20''$ (3.4 kpc) radius.  Two  faint CO streams 
stem from the western and eastern ends of the CO bar 
and extend  out in a  northeast and southwest direction. 
The southern CO stream  suggests a very loosely wound spiral arm. 
In Fig 7b. (Plate 2),  it is significant and striking  that 
the  CO streams tend to  lie along the two relatively straight 
dust lanes  in the B-V image.  
As described in $\S$~4.1, these dust lanes are 
offset towards the leading edge of the large-scale 
stellar oval  of radius $25''$ (4.3 kpc), which 
might be a  primary stellar bar. 
Prominent dust lanes along the leading edges of a stellar bar 
trace the loci of shocks which cause bar-driven gas inflow 
(e.g., Tubbs 1982; Schwarz  1984; 
Combes \& Gerin 1985;  Athanassoula 1992; 
Byrd et al 1994; Piner et al. 1995).  
It is therefore likely that the CO streams represent 
the gas still inflowing  towards the CN region 
under the action of the large-scale stellar oval.

\vspace{0.2in}
\subsection{How massive is the bar-like molecular feature?}
The mass and dynamical mass fraction of the bar-like CO feature 
are  important for differentiating between  scenarios 
for the formation of nuclear bars ($\S$ 4.8), and for 
predicting how the circumnuclear region will evolve ($\S$ 4.7).  
We estimate the  mass of molecular hydrogen (M$_{\rm H2}$)  
from the relation (Kenney \& Young 1989; Scoville \& Sanders  1987) :  

\begin{equation}
\frac {M_{\rm H2}} {(M_{\rm \sun})} = 1.1 \times 10^4 
\ (\frac {\chi}{2.8 \times 10^{20} })\ (D^2) \ (\int{S_{\rm CO}}dV)
\end{equation} 
where D is the  distance  in Mpc,  $ \int{S_{\rm CO}}dV$ 
is the integrated line flux  in Jy km s$^{-1}$, and 
$\chi$ is the CO-H$_{\rm 2}$ conversion factor 
(defined as the ratio of the beam-averaged column density of 
hydrogen to the integrated CO 
brightness temperature). Is the  Milky Way value of 
$2.8 \times 10^{20}$  H$_{\rm 2}$ cm$^{-2}$  ( K km s$^{-1}$)$^{-1}$  
appropriate for $\chi$  in a   circumnuclear starburst region, 
where molecular  gas might not be in 
virial equilibrium, and  the gas temperature T and density $ \rho $ 
can be  significantly higher than in Milky Way clouds?  
Since $\chi$ depends on  $ \sqrt \rho/T $  
(Scoville \& Sanders 1987), one might argue that 
the effects of elevated temperatures and densities will partially 
offset each other. On the other hand, multiple-line studies  and radiative 
transfer models (Wall \& Jaffe 1990; Wild et al. 1992; 
Helfer \& Blitz 1993; Aalto et al. 1995) 
have suggested that $\chi$  is lower than the Milky 
Way value by a factor of $\sim$ 3  in the centers of some starburst galaxies. 
Given the absence of multiple-line studies in NGC~2782, 
we cannot determine if $\chi$ deviates significantly 
from the  Milky Way value, and we therefore  
choose to express  many 
results in this paper  explicitly in terms of $\chi_{\rm 1}$, where 
$\chi_{\rm 1}$ = ($\chi$/$2.8 \times 10^{20}$).

The  hydrogen mass (M$_{\rm H2}$) in the central $20''$ (3.4 kpc) 
radius is  $\sim$ ($2.4 \times 10^9$ $\chi_{\rm 1}$) M$_{\tiny \odot}$.
The total mass  of molecular gas including the contribution 
of He is $\sim$ ($3.3 \times 10^9$ $\chi_{\rm 1}$) M$_{\tiny \odot}$, 
assuming a solar composition. 
About 75~\% or ($2.5 \times 10^9$  $\chi_{\rm 1}$) M$_{\tiny \odot}$
of this gas lies in the bar-like CO feature  of radius $7.5''$ (1.3 kpc). 
The  bar-like molecular feature is therefore quite massive if 
$\chi_{\rm 1}$ is not significantly less than 1.
The dynamical mass enclosed within the radius ($\sim$ $7.5''$ or 1.3 kpc)
of the nuclear stellar bar is ($8 \times 10^{9}$/ sin$^2i_{\rm co}$)
 M$_{\tiny \odot}$.  This value is derived assuming 
a flat rotational  speed  of (165 km s$^{-1}$/sin $i_{\rm co}$) 
from the CO velocity field, 
where i$_{\rm co}$ is the inclination of the circumnuclear 
molecular gas. 
The molecular gas   ($2.5 \times 10^9$  $\chi_{\rm 1}$ M$_{\tiny \odot}$) 
therefore makes up  (30~\%  sin$^2i_{\rm co}$  $\chi_{\rm 1}$ ) 
of the dynamical mass. 
With  i$_{\rm co}$ $\sim$ 30 $\deg$ ($\S$ 4.3), 
the gas mass fraction is   (7.5 $\chi_{\rm 1}$)~\% .

\subsection{Circumnuclear star formation and dust morphologies}


The H$\alpha$  and [O~III] observations presented in 
Paper I  show that  star formation in the 
disk of NGC 2782  is concentrated in the  
bar-like CO feature  of radius $7.5''$ (1.3 kpc) and  
along the ripple located at  $\sim$ $25''$ (4.3 kpc). 
We estimate the SFR in the CO bar from NIR 
data rather than optical tracers in order  to reduce 
extinction problems. Using the Br$\gamma$ luminosity  
($1.6 \times 10^{10}$ L$_{\sun}$; Puxley et al. 1990) 
in the central 10$''$ (1.7 kpc) radius , 
and assuming a case B recombination 
with an electron density $n_{e}$ of 10$^4$  cm$^{-3}$, 
a temperature of 10$^4$ K, and an extended Miller-Scalo 
IMF (Kennicutt 1983), 
we derive a  SFR of  5-6 M$_{\tiny \odot}$ yr$^{-1}$.
An alternative estimate of the SFR can be obtained 
by assuming that the total FIR luminosity 
($2 \times 10^{10}$ L$_{\sun}$; Smith 1991) 
originates predominantly from dust heated by massive stars 
in the circumnuclear region. Applying 
the method described by Hunter et al. (1986) then 
gives  a SFR of $\sim$ 4/$\beta $  M$_{\tiny \odot}$ yr$^{-1}$, 
where $\beta$ represents the 
ratio of the FIR luminosity to the total bolometric 
luminosity of the massive stars. For $\beta $ between 0.5 
and 1, the circumnuclear SFR lies  between  4 to 8  
M$_{\tiny \odot}$ yr$^{-1}$. 
Star formation is happening throughout the bar-like
CO feature, but it is mostly  concentrated 
between the CO lobes in central 200 pc radius, as 
seen in both H$\alpha$ and 5  GHz RC (Fig. 3d). 
This central starburst has a  SFR of 
3 to 6 M$_{\tiny \odot}$ yr$^{-1}$, 
comparable to the powerful archetypal starburst M82.

Fig. 8 shows the HST H$\alpha$+[N~II] image 
with a $10''$ usable   field of view. 
The HST  image reveals a wealth of details within 
features which were previously identified in 
ground-based images. 
The southern outflow `bubble' (Jogee et al. 1998)  is clearly 
resolved, while  knots  of  HII regions stand out  in the 
arc of star formation which lies north of the central starburst 
and extends  approximately east-west. 
The HST V-band (F555W) with a  $35''$ (6 kpc) field of view 
is  shown in Fig. 9a. The uniformly weighted 
($2.1'' \times 1.5''$ or $355 \times 255 $ pc)  CO  map 
is superposed as contours on the V  image in  Fig. 9b. 
Figs. 9a-b show  the northern  arc of star formation 
and reveal conspicuous  dust lanes which stem from the western 
end of the arc and extend northeast out to a 
radius of $25''$ (4.3 kpc).  
These dust lanes coincide with those  seen in the ground-based 
B-V image (Fig. 7a; Plate L2),  offset towards the leading edge of the 
large-scale stellar oval shown in Fig. 3a.  
Furthermore, very striking in Fig. 9a but not evident in 
ground-based images, are  dust lanes  running approximately east-west, 
about $4''$ (680 pc) north of the nucleus. These nuclear  dust lanes, 
are offset in a leading sense  with respect to 
the nuclear CO and stellar bars, and could represent  
shocks associated with these nuclear bars. 
The fact that these dust lanes are conspicuous to the 
north of the nucleus, but not to the south,  
supports our contention (Paper I) that the northern  
side of the inner disk is the near side.

\subsection{The nuclear CO bar in NGC~2782 compared to other systems}

While a stellar bar can be defined in terms of periodic 
families of stellar  orbits, there is no firm definition of what constitutes 
a gas bar since  gas, being  collisional and dissipative, 
does not follow closed periodic orbits, especially crossing orbits. 
In the literature, gas  distributions which are observed to be 
elongated, double peaked, or double-lobed have all been loosely 
called `gas bars'. 
It is therefore  important to  characterise how the bar-like CO feature 
in NGC 2782  differs from other  bar-like 
circumnuclear CO morphologies:

\it
(i)
\it 
Highly inclined systems: 
\rm
A high inclination as well as low resolution  CO data 
makes it ambiguous whether a bar-like  CO morphology 
is a molecular bar or a relatively axisymmetric 
gas ring viewed edge-on. The nearly edge-on 
minor mergers NGC 2146 and NGC 3079 are two typical examples.
CO observations of NGC~3079  at $4''$ (300 pc) resolution 
(Irwin \& Sofue 1992) 
show a centrally peaked elongated component  of radius $\sim$ 800 pc.
In NGC~2146, $7''$ (700 pc) resolution CO observations show 
strong CO emission from two lobes which lie $\sim$ 1 kpc apart 
(Jackson \& Ho 1988; Young et al. 1988).  
However, the  double-lobed  CO morphology in NGC 2782 is 
unlikely to be  the result of a high inclination.
The CO map in Fig. 5a  is not suggestive of a highly inclined ring, 
and  we have already presented several arguments 
for a moderate inclination (30 $\deg$) rather than 
an  edge-on inclination in $\S$~4.3.  
Hence, NGC 2782 likely has  an 
\it
intrinsically
\rm 
bar-like, non-axisymmetric gas distribution in the galactic plane.

\it
(ii) 
\rm 
\it 
Gas concentrations near ILRs in  galaxies without a  nuclear 
stellar bar :  
\rm 
A bar-like  double-peaked CO  morphology  
in the moderately inclined galaxy   NGC 6951 
has been interpreted as gas piling up near 
inner Lindblad resonances (Kenney et al. 1992). 
Several differences  demarcate NGC 2782 from NGC 6951. 
While the star formation in  NGC 6951 
peaks in a ring-like configuration near the ILRs, 
the star formation in NGC 2782 peaks sharply
\it 
in the central 200 pc, interior to the two CO lobes. 
\rm
This suggests that  in the recent past of NGC 2782, 
gas  must have reached  the central starburst 
in  the inner 200 pc. There is no published evidence for 
a nuclear stellar bar  in  NGC 6951, and 
the CO peaks lie near the turnover point of the rotation curve. 
On the other hand, NGC 2782 hosts a kpc-sized nuclear stellar bar, 
and the CO lobes 
\it
lie on the leading edges of this nuclear stellar bar, 
\rm
well inside the turnover radius ($\sim$ $6''$ or 1 kpc) of 
the rotation curve, and show bar-like streaming 
motions ($\S$ 4.2). 
This  suggests that while in NGC 6951 gas may be  piling up near 
ILRs, in NGC 2782 it  may be   experiencing further 
inflow from the  CO lobes into the central starburst, likely 
under the action  of the nuclear stellar bar.  
In fact, the formation of a nuclear stellar bar  
might be a way of converting   non-starbursts 
like NGC 3351 and NGC 6951 into centrally concentrated  
starbursts like NGC~4102, NGC~4536, NGC~3504, and 
NGC~470  (Jogee 1998; Jogee \& Kenney 1998).

\it 
(iii) 
\it 
Other  systems with   potential nuclear gas bars: 
\rm
Nuclear stellar bars observed to date (Elmegreen et al.  1996; 
Friedli et al. 1996; Jungwiert et al. 1997) 
have a radius ranging from a few hundred pc to a kpc. 
High resolution interferometric CO observations needed  to identify any 
bar-like molecular counterpart are  not available 
in many cases such as  NGC 1097, NGC 5850,  and NGC 3981. 
A similar case is   NGC 7552 where K-band and 
$H_{\rm 2}$ S(1) 2.12 $\micron$ images (Schinnerer et al. 1997) 
show an oval feature of radius $\sim$ $5''$ (500 pc).
In  other galaxies  such as   Maffei 2 
(Hurt \& Turner 1991)  and IC342 (Ishizuki et al. 1990) 
CO observations show  elongated  bar-like CO distributions in 
the inner 500 pc radius, but there is no conclusive evidence for 
a nuclear stellar bar.

In the case  of M100 (NGC~4321), both high resolution CO 
and NIR data exist, but differing model-dependent  
interpretations  have been proposed.
NIR images of  M100  (Knapen et al. 1995a)  show a stellar oval  of radius  
$4''$ (320 pc),  offset by only 4 $\deg$ from 
a large-scale stellar bar of radius $60"$ (5 kpc).
CO maps (Rand 1995; Garcia-Burillo et al. 1998) 
show two trailing molecular gas spiral arms  which extend 
from a  gas concentration  in the central $\sim$~$1.5''$ (100 pc) radius. 
Numerical models  of Garcia-Burillo et al. (1998)   suggest 
that M100 harbors a fast-rotating nuclear  stellar bar 
nested within a large-scale bar, while models  
of Knapen et al. (1995b) 
suggest that  a single large-scale bar exists.

\vspace{0.1in}
\subsection{The evolution of the  circumnuclear region}

There exists few direct observational studies of how the  
circumnuclear region evolves under the impact of a nuclear stellar bar. 
We address this issue in NGC 2782 where  we have concurrent 
observations which reveal the distribution of molecular
gas and star formation in the region of  the nuclear stellar 
bar. We emphasise that 
given the complexity of the interplay, the ensuing discussion 
should be taken in a qualitative sense.

\subsubsection{The gravitational torque 
exerted by the nuclear stellar bar}

The nuclear gas bar is offset towards the leading edge
of the nuclear stellar bar ($\S$ 4.2) .
We therefore expect the gravitational torque exerted by the 
nuclear stellar bar on the gas bar to remove angular momentum
from the gas. The net radial inflow rate cannot be directly 
measured for several reasons. The galaxy is relatively 
face-on (i $<$ 30 $\deg$) and hence the line-of-sight 
velocities  are not very  sensitive to in-plane azimuthal 
and radial velocities. 
Furthermore, along  a general position angle, the line-of-sight 
velocity is made up of a combination of in-plane 
azimuthal and radial motions, and out-of-plane vertical motions. 
It is easiest to isolate the in-plane radial component 
along the kinematic minor axis where the 
in-plane azimuthal motions are negligible.
However, the nuclear stellar bar along which one might expect radial 
gas inflow is at a  very different PA (100-110 $\deg$) from the kinematic 
minor axis  (165 $\deg$).  This orientation makes the direct 
observation of the radial component difficult. 
We therefore adopt a semi-empirical 
method  where we try to estimate the torque from the data,  
and hence derive the radial component of the velocity at different
positions.

In the ensuing discussion, we use unit position vectors 
(\bf
e$_{\bf r}$,~e$_{\bf \theta}$,~e$_{\bf z}$\rm) 
to describe the circumnuclear region of the galaxy, 
and $\Phi$ to describe the stellar potential. The unit vector 
\bf
e$_{\bf z}$
\rm
is perpendicular to the plane of the nuclear stellar bar, 
(\bf e$_{\bf r}$, e$_{\bf \theta}$\rm) are 
in the plane of the bar, and the angle $\theta$ is measured 
w.r.t. the major axis of the nuclear stellar bar. 
The torque per unit gas mass 
exerted by the nuclear stellar bar on a gas element 
at a position (r, $\theta$) is :

\begin{equation}
\bf \tau \rm = 
\frac {\partial\Phi} {\partial\theta } \ \bf e_{\bf z} \rm  - 
r \ \frac {\partial\Phi} {\partial z} \ \bf e_{\bf \theta} \rm
\end{equation}

\noindent
Gas shocked on the leading edge of the 
nuclear stellar bar loses angular momentum per unit gas 
mass (L$_{z}$) at the rate : 

\begin{equation}
\frac {dL_{z}} {dt } \ \bf e_{\bf z} \rm  \approx 
2 \ r \ \Omega_{s} \ \frac {dr} {dt}  \bf e_{\bf z} 
\end{equation} 

\noindent
where $\Omega_{s}$ is the pattern speed of the 
secondary bar. 
Since 
d
\bf
L
\rm
/dt = 
\bf 
$\tau$
\rm
, the radial  velocity is given by,

\begin{equation}
\frac {dr} {dt}   \approx
\frac {1} { \ 2 \ r \ \Omega_{s}} 
\frac {d\Phi} {d\theta} 
\end{equation}

\noindent

The expression we derived for the radial velocity dr/dt in Eq. 4 
is consistent with the one given by Quillen et al. (1995).
The term  involving  $\Phi$ in the 
above equation can be evaluated in several ways.
One method is to convolve the K-band image to 
obtain the barred stellar potential (e.g., see Quillen et al. 1995). 
This method requires detailed modeling and 
knowledge of the mass-to-light 
ratio in the region of the stellar bar. 
In the circumnuclear starburst region of NGC~2782, 
the mass-to-light ratio is uncertain. 
We therefore adopt a different, albeit more simplistic approach.
We  approximate the potential of the nuclear stellar bar 
with an analytic expression for a  barred logarithmic 
potential (Binney \& Tremaine 1987): 

\begin{equation}
\Phi(r,\theta) = 
\frac {V_{o}^{2}}{2} \ 
\rm ln \ [ \ R_{c}^{2} \ + \ r^{2}(\rm cos^{2}\theta + 
\frac {\rm sin^{2}\theta}{q^{2}}) \ ]
\end{equation} 

\noindent
In the above expression, R$_{c}$ is a core radius, 
q is the axial ratio, and V$_{o}$ is a circular speed. 
Using Eq. 4, the radial velocity  dr/dt is then :

\begin{equation}
\frac {dr} {dt}  \approx 
\frac { V_{o}^{2} r} {4 \Omega_{s}} 
\rm sin2\theta
\ [\frac { (1/q^{2} -1)} 
{R_{c}^2 + r^{2}(\rm cos^{2}\theta + \rm sin^{2}\theta/q^{2})}]
\end{equation}

\noindent
It is noteworthy that for  radii r $\gg$ R$_{c}$, the 
radial velocity 
is highly sensitive to the term (sin2$\theta$/$\Omega_s$r). 
This dependence correctly  predicts  that 
gas elements on the minor axis of the nuclear stellar 
bar  (where $\theta$ =90 $\deg$,  270 $\deg$), 
will experience no net radial inflow. 
Furthermore, nuclear gas bars which are 
offset by relatively large angles (e.g., 45 $\deg$)  with respect to 
the nuclear stellar bar major axis will be rapidly driven 
inwards. This  suggestion  is consistent with 
simulations (e.g., Friedli \& Martinet 1993), 
where the nuclear gas bar is observed 
to lead the nuclear stellar bar only by a small 
angle of order  5 to 10 $\deg$. 

In order to estimate  the radial velocity  for NGC~2782, 
the constants $\Omega_s$, V$_{o}$, and q must be evaluated. 
The pattern speed  ($\Omega_s$)  of the nuclear 
stellar bar is estimated  by assuming that the corotation  resonance 
of the nuclear stellar bar is near the radius ($\sim$ $8''$ or 1.4 kpc) 
of the  ellipticity minimum in Fig. 4. 
The condition [$\Omega$ = $\Omega_s$] , where $\Omega$ is the 
angular speed, leads to 
 $\Omega_s$ $\sim$  (123  km s$^{-1}$ kpc$^{-1}$ /sin $i_{\rm co}$).  
With  i$_{\rm co}$ $\sim$ 30 $\deg$ ($\S$ 4.3), 
the nuclear  bar pattern speed is  $\sim$ 245  km s$^{-1}$ kpc$^{-1}$, 
corresponding to a rotation period of $\sim$ $3 \times 10^7$ years.
In order to determine V$_{o}$, we note that in the limit 
r~$\gg$~R$_{c}$ and q=1, the potential $\Phi$ produces 
a flat rotation curve with speed V$_{o}$. 
For NGC~2782, we assume that the turnover velocity of 
(165 km s$^{-1}$/sin $i_{\rm co}$) given by the CO kinematics 
($\S$~4.2) is a good approximation for  V$_{o}$. 
With $i_{\rm co}$~=~ 30 $\deg$, V$_{o}$~$\sim$~330  km s$^{-1}$. 
We adopt an axial ratio q $\sim$ 0.85 since the 
ellipticity of the nuclear stellar bar in the K-band light
is 0.28, and the stellar potential will have a lower 
ellipticity since it is a convolution of the density profile. 
Near the bar end, at a radius r $\sim$ $6''$ (1 kpc),  gas elements  
located at $\theta$ $\sim$ 5-20 $\deg$  from the stellar 
bar in NGC~2782 have radial velocities  
of $\sim$ 15-45  km s$^{-1}$. It is important to realise that 
the  local radial  velocity (15-45  km s$^{-1}$) 
is not necessarily equal to the  net radial mass inflow velocity. 
This is because the  net inflow rate depends on the 
radial inflow velocity  at \it all \rm parts of the gas 
orbit,  and not only when it is on the leading edge of the 
bar.  The true average mass inflow velocity is probably lower. 
The bar-like CO feature contains  
$\sim$ $2.5 \times 10^9$   M$_{\tiny \odot}$,  with 
several clumps of mass   $10^7$ - $10^8$ M$_{\tiny \odot}$
within the inner kpc radius. 
With even a very conservative average inflow velocity of 5 km s$^{-1}$, 
for a gas element  of  $10^7$ M$_{\tiny \odot}$,  a mass inflow rate of 
1 M$_{\tiny \odot}$ yr$^{-1}$ is easily achieved. This suggests that  
gravitational torques from the  nuclear stellar bar 
may produce   large  inflow rates 
($>$ 1 M$_{\tiny \odot}$ yr$^{-1}$) of the molecular gas.

\subsubsection{Dynamical friction}

\medskip
Mechanisms other than gravity torques   can also 
be important for driving gas inwards in NGC 2782. 
Dynamical friction 
(whose timescale is $\propto$ r$^2$) becomes increasingly 
important at smaller  radii.
A body of mass M which is moving at speed v at a 
radius r, can be driven to the center by 
dynamical friction  on  a timescale (t$_{df}$) 
$\propto$ (r$^2$ v/M ln$\Lambda$), 
where ln$\Lambda$ is the Coulomb logarithm (Binney \& Tremaine 1987). 
The molecular gas bar in NGC~2782 contains several 
clumps which have sizes  between 
$2''$-$3''$, masses 
$\sim$ ($10^7$ - $10^8$) M$_{\tiny \odot}$ 
(for $\chi_{\rm 1}$~=~1), 
and are located within a radius of $4''$ (700 pc).
The dynamical mass within a radius of $4''$ 
 is $\sim$ $ 7.5 \times 10^{9}$  M$_{\tiny \odot}$ 
(for  i$_{co}$ = 30 $\deg$), and significantly exceeds 
the mass of molecular gas.  Therefore, if the gas clumps are 
self gravitating entities, they  
can experience a significant drag due to dynamical 
friction from the stellar background. 
For M $\sim$ $10^8$ M$_{\tiny \odot}$, r $\sim$ 700 pc, 
a rotational speed v $\sim$ 220 km s$^{-1}$ 
(derived from the CO velocity field for i$_{co}$ = 30 $\deg$), 
t$_{df}$ $\sim$ $ 5 \times 10^{7}$ years. 
At smaller radii, the dynamical friction 
timescale drops sharply. 
For  a conservative dynamical friction timescale of 
$\sim$ $ 5 \times 10^{7}$ years, and a 
total gas mass in the form of clumps of $\sim$  
($5 \times 10^8$  M$_{\tiny \odot}$), the average gas inflow rate 
is $\sim$ 10 M$_{\tiny \odot}$ yr$^{-1}$. 
However, this  estimate should be taken with 
caution. The sizes of the clumps ($2''$ - $3''$) are comparable  to 
the resolution ($\sim$ $2''$) of our CO observations, and 
it is therefore possible that each clump is made up of 
several smaller, less massive sub-clumps that we 
cannot presently resolve. With less massive clumps, the dynamical
friction might only become a rapid transport mechanism 
at smaller radii.

\subsubsection{Circumnuclear Evolution}

The short dynamical timescale of $\sim  4  \times 10^{6}$ 
years (Boer et al.  1992) for the starburst-driven outflow, 
and the collimated H$\alpha$ bubble morphology with a closed shell 
at the edge of the bubble (Jogee et al. 1998),  suggest that the 
outflow and starburst are probably less than  10$^{7}$ years old. 
How long can the central starburst in the inner 200 pc 
radius be sustained? 
The SFR is $\sim$  3-6 M$_{\tiny \odot}$ yr$^{-1}$ in the central starburst, 
but is relatively low  (1-2 M$_{\tiny \odot}$ yr$^{-1}$) in the CO lobes 
which contain  most of the circumnuclear  molecular gas. 
Even in the limiting case where  all of the 
$2.4 \times 10^9$  M$_{\tiny \odot}$  of molecular gas in the CO lobes  
is fed into the central starburst,  it can  only be sustained  
for $\sim$  $ 5 \times 10^8$ years.
In practice, the lifetime might be even shorter 
since some of the gas in the lobes gets converted into stars 
before reaching the starburst region, and part of the interstellar 
medium might  be blown out  by the starburst-driven wind. 
In Paper I, we estimated that the outflow 
 contains $\sim 10^{5}$  M$_{\tiny \odot}$  of 
hot and warm ionised gas, and possibly 
$\sim 2 \times 10^{7}$  M$_{\tiny \odot}$ of cold gas.

What about the  nuclear stellar bar?
Simulations show that as a nuclear stellar bar drives 
gas into the inner few hundred pc, it may 
be  destroyed and even  evolve into a 
triaxial bulge  (e.g., Friedli \& Martinet 1993). 
Any existing large-scale primary stellar bar may be  weakened or 
completely destroyed  (Hasan \& Norman 1990; 
Friedli \& Benz  1995; Norman, Sellwood, \& Hasan 1996). 
The destruction of the stellar bars is caused by  the development of 
chaotic orbits  and an increased  mass concentration 
in the region of  x$_2$  stellar orbits 
which are  perpendicular to the bar. 
Typically,  bar dissolution occurs 
if the  mass concentration in the central 200 pc radius 
is $\ge$  1-2~\% of the total stellar mass in the disk. 
The  nuclear stellar bar dissolves over  a few  rotation 
periods, typically a few  $\times 10^8$ years 
(Friedli \& Martinet 1993; Combes 1994).  
In NGC~2782, for a standard CO-H$_{\rm 2}$ conversion factor, 
the gas mass ($2.5 \times 10^9$  M$_{\tiny \odot}$)  in the CO lobes 
is  $\sim$~1~\%  of the total stellar mass in 
the optical disk of radius  1' (10 kpc). 
Thus,  if  most of this  gas is  driven into the inner 200 pc 
radius, the nuclear stellar bar and the large-scale stellar oval 
can be rapidly destroyed.  
Nuclear stellar bars are also expected to be short-lived 
as a result of the rapid transfer of angular 
momentum from the bar to the galactic halo via dynamical 
friction, on timescales of a few rotation periods 
(Weinberg 1985), or a few $ \times 10^8$ years.

\subsection{Forming nuclear bars : Comparison with simulations and data}


Simulations have suggested different ways of forming nuclear 
stellar and gas bars.
Some studies have focused on  gas-rich systems where the 
circumnuclear gas mass fraction is between 20 to 50 \% 
(Shlosman et al. 1989; Heller \& Shlosman 1994). 
Typically, a circumnuclear gas-rich disk forms, it 
becomes bar-unstable, and the nuclear gas bar rapidly fragments.
However, nuclear stellar bars do not form in these 
simulations. This characteristic is consistent with 
the simulations and analytical work of Shlosman \& Noguchi (1993) 
who find that for a  disk with 
a gas mass fraction well above 10~\%, the gas becomes  clumpy, and 
dynamical fraction between the gas clumps and the 
sea of stars  heats up the stellar disk over  several rotation periods. 
This  results in a hot stellar disk which is stable against 
bar instabilities.

More relevant for NGC 2782  might be the  simulations 
of Friedli \& Martinet (1993) or   Combes (1994), which produce 
nuclear stellar bars in moderately gas-rich circumnuclear regions. 
A  nuclear stellar bar which is rotating faster than the large-scale 
primary stellar bar can form if the mass 
density contrast between the inner  and outer 
regions of the primary bar  is large enough.  
Molecular gas is critical for the decoupling : 
it helps to increase  the central mass 
concentration and it loses angular momentum to the 
nuclear stellar bar, thereby helping it to  
maintain a large pattern speed. 
The molecular gas distribution becomes bar-like and this 
nuclear gas bar  leads the nuclear stellar bar by 
a small angle, and 
can make up $\sim$  10~\%  of the  dynamical mass. 
NGC~2782 has many  properties consistent with these 
simulations: it hosts a clumpy bar-like CO  feature whose  
radius  (7.5$''$ or 1.3 kpc) is comparable  to that of 
the nuclear stellar bar; the CO peaks 
lie on the leading edges of the nuclear stellar bar 
and show weak bar-like streaming motions;  
the molecular gas makes up   $\sim$   7.5 \% 
of the dynamical mass within a $7.5''$  (1.3 kpc) radius 
for a standard CO-H$_{\rm 2}$ conversion factor. 
However, one feature not so evident in these simulations, 
but particularly striking in NGC 2782, is the clumpiness of the 
molecular gas bar (Fig. 5a). 
If these clumps are self-gravitating entities, then dynamical 
friction  ($\S$ 4.7.2) might be at least as important 
as gravitational torques in driving the gas towards the center.

Simulations  by Shaw et al. (1993) also 
produce a nuclear stellar bar, but it is still 
coupled to the primary stellar bar. 
In this scenario, a gas ring near the ILRs is phase-shifted in 
\it 
a leading sense 
\rm 
with respect to the large-scale primary stellar  bar, 
and the gravitational perturbation of the gas on the 
stellar orbits produces isophotal twists.
The distorted isophotes can give the appearance 
of a nuclear stellar bar  which 
\it  
always  leads 
\rm
the primary bar. 
This contrasts with  the 
dynamically decoupled nuclear stellar bar and gaseous bar 
which can lead  the primary bar part of the time and 
\it
trail 
\rm 
it at other times.
In the case of NGC 2782, the nuclear stellar and gas bars 
are  nested in a large-scale stellar oval of 
radius $25''$  (4.3 kpc) which could be primary stellar bar 
($\S$ 4.1). 
The clumpy nuclear gas bar  with a  PA between 70 to 85 $\deg$, 
trails the large-scale stellar oval  whose PA lies 
between 20 to 30 $\deg$.
Hence, if the large-scale stellar oval is a primary bar,  then 
NGC 2782 supports the scenarios to form 
decoupled nuclear bars.

However, the simulations discussed so far deal with isolated 
galaxies where 
a spontaneously-induced primary bar was the main way 
to drive gas from the outer disk to the inner kpc.  
In the case of NGC 2782, which has experienced a recent 
interaction or merger (Smith 1994), 
such gas inflows can result from the interaction itself, namely 
direct tidal torques  (Hernquist \& Mihos 1995), 
as well as gravitational torques from tidally-induced 
large-scale stellar bars 
(Noguchi 1988; Combes et al. 1990; Barnes \& Hernquist 1996). 
Once gas reaches the circumnuclear region, the formation of 
nuclear bars may ensue in a fashion qualitatively similar to 
the scenarios for isolated galaxies. 
Another alternative for interacting galaxies 
is that a  nuclear stellar bar may form 
by the accretion of a rapidly rotating gas-rich companion. 
In fact, decoupled nuclear stellar disks/bars are 
quite common in  HST  images  
(Barth et al. 1995), and in spectroscopic studies by 
Rubin at al. (1997) who find that $\sim$ 20~\% of a sample of 
80 Virgo cluster galaxies host a decoupled disk or ring 
within a 500 pc radius.

We also suggest that NGC 2782 would not harbor a 
nuclear stellar bar if  it had undergone a major merger. 
Violent relaxation, which operate in major mergers, tends 
to eradicate non-axisymmetric structures and  produce  
an  r$^{1/4}$ stellar surface brightness profile.   
The large gas  inflow rates and  large central  gas 
concentrations typical of major mergers  can  rapidly 
destroy  primary and nuclear stellar bars, as described in 
$\S$ 4.7.3. 
The  large gas mass fraction also leads to the formation of gas clumps 
which heat the stellar component  via dynamical friction, 
thereby making it stable   against bar  formation 
(Shlosman \& Noguchi 1993). 
For instance, the major merger NGC~7252  has an r$^{1/4}$  
K-band profile (Schweizer 1982) and the molecular gas 
mass fraction  exceeds 30 $\%$ in the central kpc. 
In contrast, NGC 2782 has a molecular gas mass fraction of 
$\sim$  8~\%  in the inner kpc radius.
It has  properties inconsistent with a major merger: 
its optical disk can be fitted with an exponential surface  
brightness profile, rather than  an r$^{1/4}$ profile 
between  radii of $\sim$ $30''$ to $60''$; 
its large-scale HI properties  have been modeled 
with an intermediate (1:4) mass ratio interaction or merger 
by Smith (1994). Furthermore, our  R-band image overlaid 
on  the HI map in Fig. 10b  shows that to the west, 
the stellar tail is shorter than 
the 50 kpc long HI tail, while to the east the opposite is true.
This opposite offset in tails on different  sides of a galaxy 
is not observed in major mergers such as NGC 7552.

\section{Summary and conclusions}

We have presented a study of the 
peculiar interacting starburst galaxy NGC~2782 (Arp 215) 
based on high resolution ($2''$) interferometric OVRO CO (J=1-$>$0) 
data, optical BVR and H$\alpha$ observations from the WIYN telescope, 
along with available NIR images, a 5 GHz RC map, and 
archival HST images. 
The optical and NIR data reveal a nuclear stellar bar of radius 
$\sim$ $7.5''$ (1.3 kpc)  nested within a large-scale 
stellar oval of radius $25''$ (4.3 kpc), whose dust lane 
morphology suggest it might be a large-scale stellar bar. 
The CO maps show  a clumpy, double-peaked  bar-like CO feature which 
is offset in a leading sense with respect to  
the  nuclear stellar bar, and has a similar radius (1.3 kpc).
The CO bar contains $2.5 \times 10^9$ M$_{\tiny \odot}$ 
of  molecular gas,  assuming a standard 
CO-H$_{\rm 2}$ conversion factor and solar metallicity. 
The molecular gas  makes up  $\sim$~8~\%  of the  
total dynamical mass within 
the  radius  (1.3 kpc) of the nuclear stellar bar. 
Within the CO bar, the CO  emission  peaks away from the nucleus,
in two  clumpy lobes lying  on opposite sides 
of the nucleus, separated by $\sim$ $6''$ (1 kpc). 
The molecular gas bar is forming stars at a rate of 
4 to 8  M$_{\tiny \odot}$ yr$^{-1}$. Most of the star formation is 
centrally concentrated between the two CO lobes, in 
the inner 200 pc radius  where the SFR is 3 to 6  M$_{\tiny \odot}$ yr$^{-1}$. 
This intense central activity, comparable to that in the prototpyical 
starburst M82,  suggests that molecular  gas fuel
must have been recently present in the  inner 200 pc radius. 
The two CO lobes lie on the leading edges of the nuclear 
stellar bar where gas is expected to lose angular momentum 
and flow inwards  as  a result of gravitational torques. 
Furthermore, while  circular motions dominate the  CO velocity field, 
the CO lobes  show weak bar-like streaming 
motions suggestive of gas inflow. 
We estimate semi-analytically the  gravitational torque by 
the nuclear stellar bar on the gas in the lobes, and our results suggest 
large gas inflow rates ($> $ 1  M$_{\tiny \odot}$ yr$^{-1}$).

While there is mounting evidence  from  NIR  surveys 
that many galaxies have nuclear stellar bars, 
the observations presented here  are amongst the first ones to 
catch a nuclear stellar bar `red-handed' in the act of feeding
molecular gas into an M82-class central starburst.
These observations are  consistent with some 
theories  and simulations which suggest that nuclear bars 
provide a  powerful way to drive  gas  closer to the  center 
to fuel a central starburst/AGN. 
In addition, we note that several massive clumps 
($10^7$ - $10^8$ M$_{\tiny \odot}$) 
are present in the inner 400 pc radius and dynamical friction might 
produce further gas inflow. 
We suggest that  as a result of star formation and gas inflow, 
the central activity, the nuclear gas bar, and 
perhaps even the nuclear stellar bar, 
will be very short-lived and  likely disappear 
within $ \rm 5 \times 10^8$ years.
The  short lifetime  suggests that 
nuclear bars can cause a rapid increase in central 
mass concentration,  trigger intense short-lived central activity, 
and play an important part in circumnuclear  evolution.

\section{Acknowledgments}
Support for this  work was generously provided by an 
AAUWEF Fellowship, 
NSF grant  AST 96-13717, a grant from the K. T. and E. L. Norris Foundation, 
a Grant-in-Aid of Research 
from Sigma Xi (The  Scientific Research Society), and 
a Zonta International Amelia Earhart Fellowship.
We thank C. Engelbracht for generously giving us the NIR images of 
NGC~2782 prior to publication, 
and D. J. Saikia and A. Pedlar for kindly providing the 
5 GHz RC maps. We thank  Richard B. Larson, F. Combes, J. Sellwood, 
and the participants of the 1996 Summer school on 
``Starbursts: Triggers, Nature and Evolution'' at Les Houches 
for useful discussions.
We thank Jeannette Barnes at NOAO for helping in the production 
of solitaires.

\clearpage
\bf
(Plate 1)--
\bf
The optical morphology: 
\rm
\noindent
The WIYN  R-band image of NGC~2782 with a  6.5$'$ (66 kpc) field of view 
reveals  an optical disk which is bracketed by two stellar tails. 
The eastern stellar plume extends to a radius of $\sim$ 
$2.7'$ (27 kpc), while the fainter, diffuse western tail stretches 
out to  $\sim$ $4'$ (40 kpc). 
Except for three ripples at radii of 25$''$, 45$''$ and 60$''$, 
the optical disk is relatively undisturbed within a  radius 
of $\sim$~$1'$ (10 kpc).

\vspace*{0.2in}
\bf
Fig. 2 --
\bf
CO (J=1-$>$0) channel maps: 
\rm
The naturally weighted channels show emission over V$_{lsr}$ = 
2364  to 2728 km s$^{-1}$ .
The velocity resolution is 10.4 km s$^{-1}$.
The contour levels are -5, -3, 3, 5, 8, 11
times the noise level (11 mJy per beam). The sythesized beam
is shown in the top panel. The cross marks the 5 GHz radio
continuum peak.

\vspace*{0.2in}
\bf
Fig. 3 --- 
\bf
The nuclear  gas and stellar bars nested in a large-scale oval 
feature: 
\bf
(a)
\rm 
The I image  with a $3'$ (30 kpc) field of view shows 
a relatively circular optical disk which has a radius 
of $\sim$ $1'$ (10 kpc). 
A weak oval feature of radius  $25''$ (4.3 kpc), at 
a position angle (PA) of $\sim$ 20 $\deg$ is 
present in the disk. It  is flanked by two 
relatively straight dust lanes which are visible in the B and B-V images 
(Figs. 7a \& 7b, Plate L2). 
The locations  of the  dust lanes are shown as solid lines here.
The oval feature might be a weak primary stellar bar.  
\bf
(b)
\rm
K  contours (Engelbracht et al. 1997) on  
high resolution ($2.1'' \times 1.5''$) CO (greyscale) for 
a $1.5'$ (15 kpc) field of view. 
The dust lanes are again shown as solid lines. 
Within the large-scale  oval of radius  $25''$ (4.3 kpc),
is nested a  gas-rich nuclear stellar bar 
which has a PA of 100 $\deg$ and a radius of $\sim$ $7.5''$ (1.3 kpc).
\bf
(c) 
\rm
K contours on CO greyscale for the central 
$20''$ (3.4 kpc) field of view only. 
The nuclear CO bar has a similar extent as the nuclear 
stellar bar and leads it by a small angle. 
\bf
(d) 
\rm
5 GHz RC  (contours, Saikia et al. 1994) 
on the high resolution CO (greyscale) for a
central $20''$ (3.4 kpc) field of view.  
The molecular gas bar is associated with star formation 
whose intensity peaks in the inner 200 pc of the bar. 
The northern and southern RC bubbles are associated with 
the starburst outflow. The two CO spurs, labelled O1 and O2, 
lie inside the  outflow bubbles and are elongated along the 
CO kinematic axis.

\vspace*{0.2in}
\bf
Fig. 4 --- 
\bf
Isophotal analysis of the K  and I images: 
\rm 
The radial profiles of surface brightness, ellipticity  and 
position angle of the azimuthally-averaged K and I  light 
are shown. 
The profiles reveal a nuclear stellar bar 
which has  a maximum ellipticity of 0.28, a 
PA of $\sim$ 100 $\deg$,  and a radius of $\sim$ $7.5''$ (1.3 kpc).
The ellipticity falls to a minimum at a radius of 
8$''$, and then smoothly  rises to $\sim$ 0.26 
at a radius of $\sim$~25$''$, while the position angle twists gradually 
from a PA of   0 $\deg$  to  $\sim$ 20 $\deg$.  It is 
possible, but not certain, that 
the second smooth rise in ellipticity 
is indicative of a large-scale primary stellar bar. 
Further out, the signal-to noise 
in the K-band image is very low and the I-band image  is used.
In this image, the ellipticity  appears to fall 
again and it reaches values below  0.1  in the outer disk 
at radii beyond $50''$ (8.5 kpc).

\vspace*{0.2in}
\bf
Fig. 5 --- 
\bf
The CO intensity and velocity fields:
\bf
(a) 
\rm
Uniformly weighted  ($2.1'' \times 1.5''$) CO total intensity map 
of the central  $20''$ (3.4 kpc). 
Most of the circumnuclear CO in NGC~2782  lies within a
radius of $8''$ (1.4 kpc), 
and has a bar-like distribution. The CO emission peaks 
in two  extended clumpy lobes which lie on opposite sides of 
the nucleus, separated by $\sim$ 1 kpc.
The CO bar leads the nuclear stellar bar whose PA is marked.
The two faint CO spurs, labelled O1 and O2, 
are elongated along the kinematic minor axis. 
The contour levels are 5, 10, 20,...100 \% of the peak 
flux. The cross marks the 5 GHz RC peak.
\bf
(b) 
\rm
Uniformly weighted ($2.1'' \times 1.5''$) CO velocity 
field (moment 1, contours) on the CO 
total intensity map. Although the gas in the central $8''$ 
(1.4 kpc) has predominantly circular motions, 
there are weak bar-like streaming motions in the CO lobes, 
and non-circular motions in the  two CO spurs.

\vspace*{0.2in}
\bf
Fig. 6 --- 
\bf
The CO kinematics: 
\rm
\bf
(a)
\rm 
The spatial velocity plot along the major  axis (75 $\deg$) 
with a width of $\pm$ $1.0''$.  The systemic velocity of  
2555 km s$^{-1}$ has been subtracted from velocities on the 
y-axis.
Complex  kinematics indicative of non-circular motions 
are present  $1''$ to $4''$ along  
the eastern and western sides of the  major axis. 
The complex   gas kinematics are associated  with  the western 
and eastern  CO peaks. The features marked as A1, A2, A3, and A4
delineate the large linewidths and  complex kinematics at the position of 
these peaks. Contour levels are 20, 25, 30, 40, 50, 60, 70, 80, 90, 
100 \% of the peak flux (73 mJy per beam). 
The cross in Figs. 6a-b marks the same position  as in Figure 5 at 
(09 10 53.65, +40 19 15.3), where the systemic velocity is 2555 
km s$^{-1}$.
\bf
(b)
\rm
The spatial velocity plot along the CO kinematic minor 
axis with a width of $\pm$ $0.5''$. 
We can see weakly emitting gas whose velocities deviate 
from the systemic velocity by $\sim$ 50 km s$^{-1}$ near O1, and by 
-30 to -50 km s$^{-1}$ near O2. 
This indicates non-circular motions  resulting from 
vertical outflow out of the disk plane  
or/and  radial gas inflow in the disk plane.
Contour levels are 2.5, 3, 3.5, 4, 4.5, 5, 6, 7 times the 
noise level (11 mJy per beam). 

\vspace*{0.2in}
\bf
Fig. 7 (Plate 2) --- 
\rm
\bf
The CO distribution in relation to the dust lanes: 
\bf 
(a)
\rm 
The  WIYN B image with a 1.6$'$ (16.3 kpc) field of view 
shows two dust lanes which are offset from the nucleus  
and extend  out to a radius of $\sim$ $25''$ (4.3 kpc). 
The two dust lanes are relatively straight and parallel to each other. 
Their appearance is similar to the dust lanes
observed along the leading edges of bars in many spiral galaxies. 
\bf
(b) 
\rm
Naturally weighted ($2.6'' \times 2.0''$)  CO intensity 
(moment 0, contours) map on the WIYN  B-V image (greyscale)
for  a $1.6'$ (16.3 kpc) field of view. 
The contour levels are 10, 20, ...100 \% of the 
of the peak  flux  (12.46 Jy km s$^{-1}$ per beam).
Notice that the dust lanes bracket the CO bar, 
and the two CO peaks  lie at 30-40 $\deg$, rather than 90 $\deg$, 
to the dust lanes. Further out, fainter streams of 
CO emission lie along the dust lanes. The southern CO stream 
suggests a loosely wound trailing spiral arm.

\bf
Fig. 8  --- 
\rm
The HST H$\alpha$+[N~II] image with  a  $7''$ (1.2 kpc)  field  of view  
shows the  central starburst, highlights 
knots  of   HII regions in the northern arc of star formation running 
approximately east-west, and  resolves 
the southern starburst-driven outflow `bubble' (Jogee et al. 1998).

\vspace*{0.2in}
\bf
Fig. 9 
\bf
(a)
\rm 
Top:   HST V-band (F555W) image with a  $35''$ (6 kpc) field of view. 
\bf
(b)
\rm
Bottom:  The uniformly weighted 
($2.1'' \times 1.5''$ or $355 \times 255 $ pc)  OVRO CO map 
superposed on the HST V image. 
These images show the 
northern arc of star formation running approximately east-west 
in the inner kpc radius, and a  set of  dust lanes extending 
northeast beyond  a radius of $20''$ (3.4 kpc). 
These dust lanes coincide with those 
seen in  the ground-based B-V image  and are offset towards the 
leading edge of the large-scale stellar oval. 
About $4''$ (680 pc) north of the nucleus, another set of  striking  
dust lanes  run approximately east-west, and  are  offset in a leading 
sense with respect to the nuclear bar-like CO feature.

\vspace*{0.2in}
 \bf
Fig.10  --- 
\bf
Comparison of the HI and stellar distributions:
\rm
\bf
(a) 
\rm 
The HI map (Smith 1994) of NGC~2782 with a $6.8'$ 
(69 kpc) field of view reveals a 
blueshifted western tail, a redshifted eastern 
tail, and an HI disk. 
Levels plotted are 10, 20,....,100 \% 
of the peak  flux of 0.34 Jy km s$^{-1}$ per beam.
The western tail  is $5'$ (50 kpc) long and 
contains 40 \% of the total HI mass of the galaxy.
\bf
(b)
\rm 
Contours of the smoothed R image on the HI map. 
The eastern might be 
\it 
stellar plume  
\rm 
extending beyond the $2'$ (20 kpc) long eastern 
\it 
HI tail. 
\rm 
In contrast,  the western diffuse 
\it
stellar tail 
\rm
seems shorter than the western 
\it
HI tail 
\rm
by $\sim$ $1'$ (10 kpc).
The cross marks the 5 GHz RC peak.


\begin{references}

Aalto, S., Booth, R. S., Black, J. H., \& Johansson, L. E. B. 
1995, A\&A, 300, 369.

 
Athanassoula, E. 1992, MNRAS, 259, 345


Balick, B., \& Heckman, T. 1981, A\&A, 96, 271

Balzano, V. A. 1983, ApJ, 268, 602


Barnes, J. E., \& Hernquist, L. 1996, ApJ, 471, 115

Barth, A. J., Ho L. C., Fillipenko, A. V., Sargent, W. L. 1995, 
AJ, 110

Binney, J.,  \& Tremaine, S. 1987,  Galactic Dynamics, 
ed. Ostriker, J. P. (Princeton, N. J.: Princeton University Press) 

Boer, B., Schulz, H., \& Keel, W. C. 1992, A\&A, 260, 67

Buta, R., \& Crocker, D. A. 1993, AJ, 105, 1344 

Bryant, P. M. \& Scoville, N. Z. 1999, AJ, 117, in press 

Byrd, G., Rautiainen, P., Salo, H., Buta, R., \& Crocher, D. A. 1994, 
AJ, 108, 476

Combes, F., \& Gerin, M. 1985, A\&A, 150, 327

Combes, F., Debasch, F., Friedli, D., \& Pfenniger, D. 1990, A\&A, 233, 82


Combes, F. 1994, in Mass-Transfer Induced Activity in Galaxies, 
ed. Shlosman, I. (Cambridge University Press), 170 



de Vaucouleurs, G., de Vaucouleurs, A., Corwin Jr., H. G., 
Buta, R. J., Paturel, G., \& Fouque, P. 1991, Third Reference Catalogue 
of Bright Galaxies  (New York: Springer) (RC3)

de Vaucouleurs, A., \& Longo, G. 1988,  ``Catalogue of Visual and Infrared 
Photometry of Galaxies from 0.5 micrometer to 10 micrometer (1961-1985)'', 
University of Texas Monographs in Astronomy (Austin: University of Texas)


Devereux, N. A. 1989, ApJ, 346, 126


Elmegreen, D. M., Elmegreen, B. G., Chromey, F. R., Hasselbacher, D. A., 
\& Bissell B. A. 1996, AJ, 111, 1880

Engelbracht, C. W., Rieke, M. J., \& Rieke, G. H. 1999, in preparation

Evans, A. S., Mazzarella, J. M., Surace, J. A., \& Sanders, D. B. 1999 
ApJ, 511, 731

Evans, I. N., Koratkar, A. P., Storchi-Bergmann, T., 
Kirkpatrick, H., Heckman, T. M., \& Wilson, S. A. 1996, ApJS, 
185, 93

Evans, A. S., Surace, J. A., \& Mazzarella, J. M. 1999, in preparation


Forbes, D. A., Ward, M. J., DePoy, D. L., Boisson, C., 
\& Smith, M. S. 1992, MNRAS, 254, 519


Friedli, D., \& Martinet, L. 1993, A\&A, 277, 27

Friedli, D. \& Benz, W. 1995, A\&A, 301, 649

Friedli, D., Wozniak, H., Rieke, M., \& Bratschi, P. 1996, 
A\&AS, 118, 461


Garcia-Burillo, Sempere, M. J., Combes, F., \& Neri, R. 1998, A\&A,333, 864


Guiricin, G., Tamburni, L., Mardirossin, F., Mezzetti, M., 
\& Monaco, P. 1994, ApJ, 427, 202

Hasan, H., \& Norman, C. 1990, ApJ, 361, 69



Helfer, T. T., \& Blitz, L., 1993, ApJ, 419, 86

Heller, C. H., \& Shlosman, I. 1994, ApJ, 424, 84

Hernquist, L., \& Mihos, J. C. 1995, ApJ, 448, 41

Hodge, P. W., \& Kennicutt, R. C. 1983, ApJ, 268, L75


Hunter, D. A., Gillett, F. C., Gallagher, J. S., Rice, W. L., \& 
Low, F. J. 1986, ApJ, 303, 171.


Hurt, R. L., \& Turner, J. L. 1991, ApJ, 377, 434



Irwin, J. A, \& Sofue, Y. 1992, ApJ, 396, L75

Ishizuki, S., Kawabe, R., Ishiguro, M., Okumura, S. K., \& 
Morita, K.-I., 1990, Nature, 344, 224

Ishuzuki, S. 1994, in Proceedings of IAU Colloqium  140, 
Astronomy with Millimeter and Submillimeter Wave Interferometry, 
eds. Ishugiro, M., \& Welch, J. M.  (ASP Conference Series), 292

Jackson, J. M.,  \& Ho, P. T. P.  1988, ApJ, 324, L5

Jogee, S., Kenney, J. D. P. \& Smith B. J., 1998, 
ApJL, 494, L185  (\bf Paper I \rm)

Jogee, S. 1998, Ph. D.  Thesis, Yale University.

Jogee, S., \& Kenney, J. D. P. 1998, Proceedings of IAU Symposium No. 184, 
The Central Regions of the Galaxy and Galaxies,  
ed. Y. Sofue  (Dordrecht: Kluwer Academic Publishers)

Jungwiert, B., Combes, F., \& Axon, D. J. 1997, A\&AS, 125, 479

Kenney, J. D. P., \& Young, J. S. 1989, ApJ, 344, 171.

Kenney, J. D. P., Wilson, C. D., Scoville, N. Z., Devereux, N. A., 
\& Young, J. S. 1992, ApJ, 395, 179


Kennicutt, R. C., Jr. 1983, ApJ, 272, 54

Kennicutt, R. C, Jr., Keel, W. C., \& Blaha, C. A. 1989, AJ, 97, 1022


Kinney, A. L., Bregman, J. N., Huggins, P. T., 
Glassgold, A. E., \& Cohen, R. D. 1984, PASP, 96, 398

Knapen, J. H., Beckman, J. E., Shlosman, I., Peletier, R. F., 
Heller, C. H., \& de Jong, R S.  1995a,  ApJL, 443, L73.


Knapen, J. H., Beckman, J. E., Heller, C. H., Shlosman, I., 
\& De Jong, R. S. 1995b, ApJ, 454, 623

Mihalas, D., \& Binney, J. 1981,  Galactic Astronomy (New York: W. H. 
Freeman and Company)



Mihos, J. C., \& Hernquist, L. 1994, ApJ, 425, L13


Negroponte, J.,  \& White, S. D. M. 1983, MNRAS, 205, 1009

Noguchi , M. 1988, A\&A, 203, 259

Norman, C. A., Sellwood, J. A., \& Hasan, H. 1996, ApJ, 
462, 114.


Padin, S., Scott, S. L., Woody, D. P., Scoville, N. Z., Seling, 
T. V., Finch, R. P., Giovanine, C. J., \& Lawrence, R. P. 1991, 
PASP, 103, 461


Pierce, J. M., \& Tully, R. B. 1988, ApJ, 330, 579

Piner, B., Glenn, S., James, M., \& Teuben, P. J. 1995, ApJ, 449, 508


Pompea, S. M., \& Rieke, G. M. 1990, ApJ, 356, 416

Puxley, P. J., Hawarden, T. G., \& Mountain, C. M. 1990, 
ApJ, 364, 77.

Quillen, A. C., Frogel, J. A.,  Kuchinski, L. E., Terndrup, 
D. M. 1995, AJ, 110, 156.


Rand, R. J. 1995, AJ, 109, 244 


Rubin, V. C., Burstein, D., Ford, Jr., W. K., \& 
Thonnard, N. 1985, ApJ, 289, 81

Rubin, V. C., Kenney, J. D. P., Young, J. S. 1997, AJ, 113, 1250

Saikia, D. J., Pedlar, A., Unger, S. W., \& Axon, D. J. 
1994, MNRAS, 270, 46

Sakka, K., Oka., S., \& Wakamatsu, K. 1973, PASJ, 25, 153


Sandage, A. 1961, The Hubble Atlas of Galaxies 
(Washington D. C.: Carnegie Institute of Washington)

Sandage, A. \& Tammann, G. A. 1981, A revised Shapley Ames Catalog 
of Bright Galaxies (Washington D. C.: Carnegie Institute of Washington) 

Sandage, A. \& Bedke, J., 1994, The Carnegie Atlas of Galaxies 
(Washington D. C.: Carnegie Institute of Washington)

Schinnerer, E., Eckart, A., Quirrenbach, A., Boker, T., 
Tacconi-Garman, L. E., Krabbe, A., \& Sternberg, A. 1997, ApJ, 488, 174


Schwarz, M. P.  1984, MNRAS, 221, 195

Schweizer, F. 1982, ApJ, 252, 455


Scoville, N. Z., Yun, M. S., Windhorst, R. A., Keel W. C.
, \& Armus, L 1997, ApJ, 485, L21

Scoville, N. Z., Yun, M. S., Clemens, D. P., Sanders, D. B., \& 
Waller, W. H. 1987, ApJS, 63, 821

Scoville, N. Z., \& Sanders, D. B. 1987, in Interstellar 
Processes, ed. D. J. Hollenbach \&  H. A. Thronson 
(Dordrecht: Reidel),  21


Scoville, N. Z., Carlstrom, J. E., Chandler, C. J.,  Phillips, 
J. A., Scott, S. L., Tilanus, R. P. J., \& Wang, Z. 1993, 
PASP, 105, 1982

Shaw, M. A., Combes, F., Axon, D. J., Wright, G. S., 1993, 
A\&A, 273, 31

Shlosman, I., Frank, J., \& Begelman, M. C. 1989, Nature, 338, 45

Shlosman, I., \& Noguchi, M. 1993, ApJ, 414, 474

Simkin, S. M., Su, H. J., Schwarz, M. P. 1980, ApJ, 237, 404

Smith, B. J. 1994, AJ, 107, 1695

Smith, B. J. 1991, ApJ, 378, 39

Smith, B. J., Curtis, S., Kenney, J. D. P., \& Jogee, S. 1999, 
AJ, 117, 1237

Teuben, P. J., Sanders, R. H., Atherton, P. D., 
van Albada, G. D. 1986, MNRAS, 221, 1
 
Tubb, A. D. 1982, ApJ, 255, 458  

Wall, W. F., \& Jaffe, D. T. 1990, ApJ, 361, L45

Weinberg, M. D. 1985, MNRAS, 213, 451.

Wild, W., Harris, A. I., Eckart, A., Genzel, R., Graf, U. U., Jackson, J. M.,
 Russell, A. P. G., \& Stutzki, J. 1992, A\&A, 265, 447


Wozniak H., Friedli D.,
Martinet L., Martin P., Bratschi P., 1995, A\&AS 111, 115

Young, J. S. et al. 1995, ApJS, 98, 219


Young, J. S., Claussen, M. J., Kleinmann, G. S., Rubin, V. C., 
\& Scoville, N. Z. 1988, ApJ, 331, L81

Yun, M. S., \&  Scoville, N. Z. 1995, ApJ, 451, L45
\end{references}
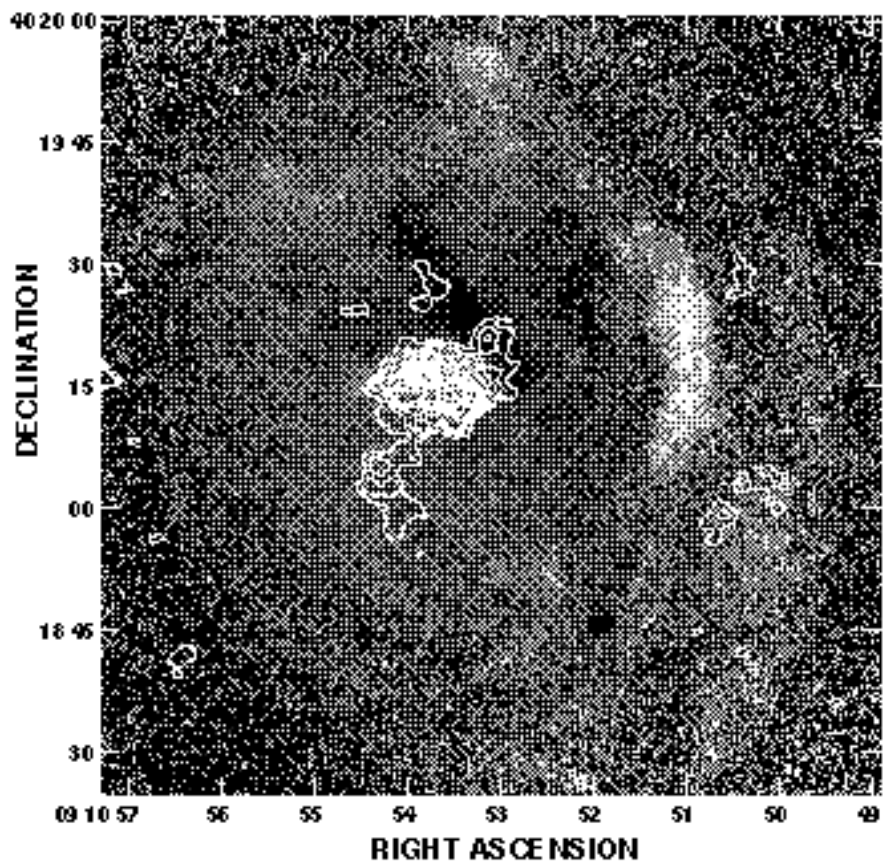
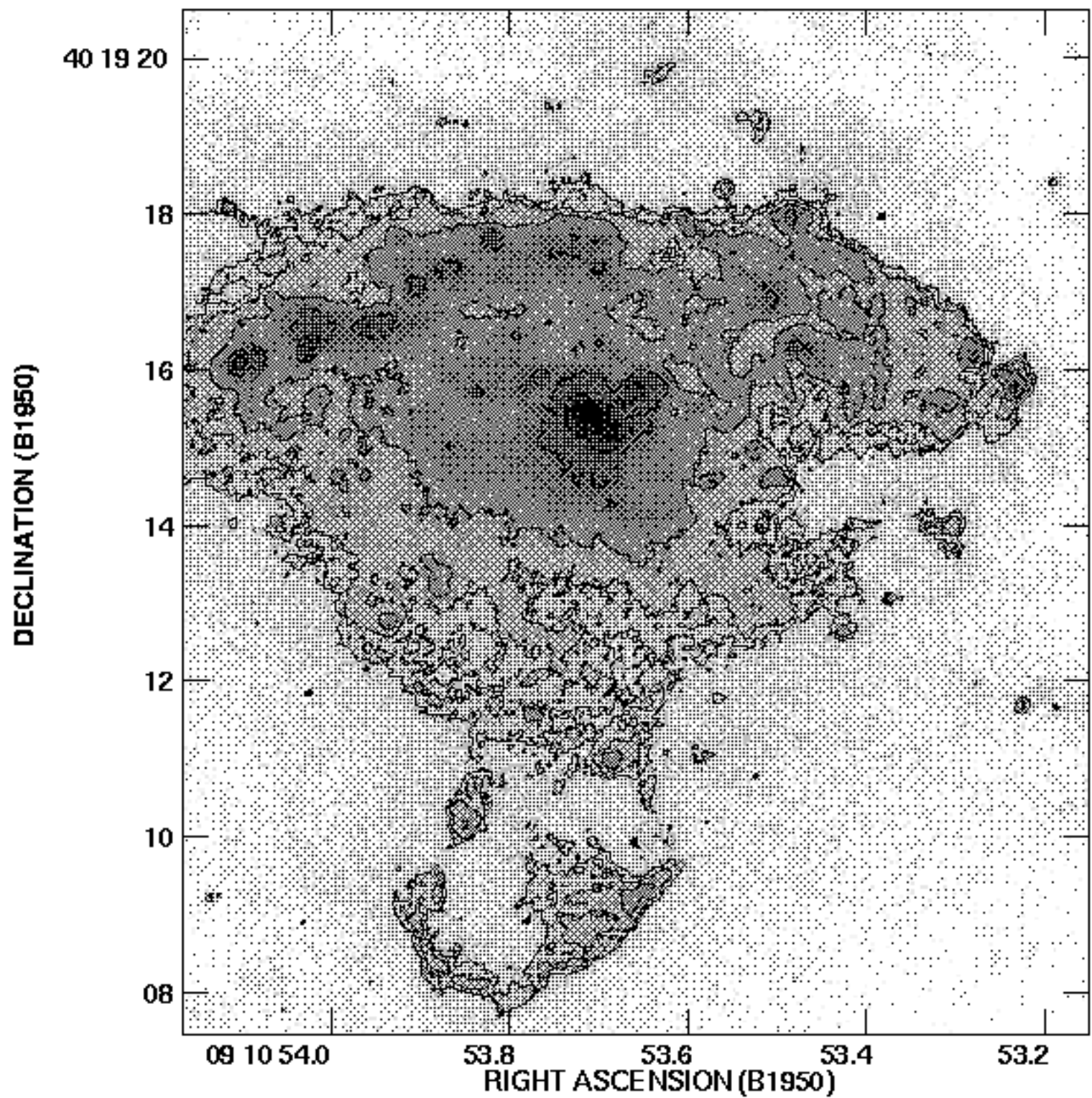
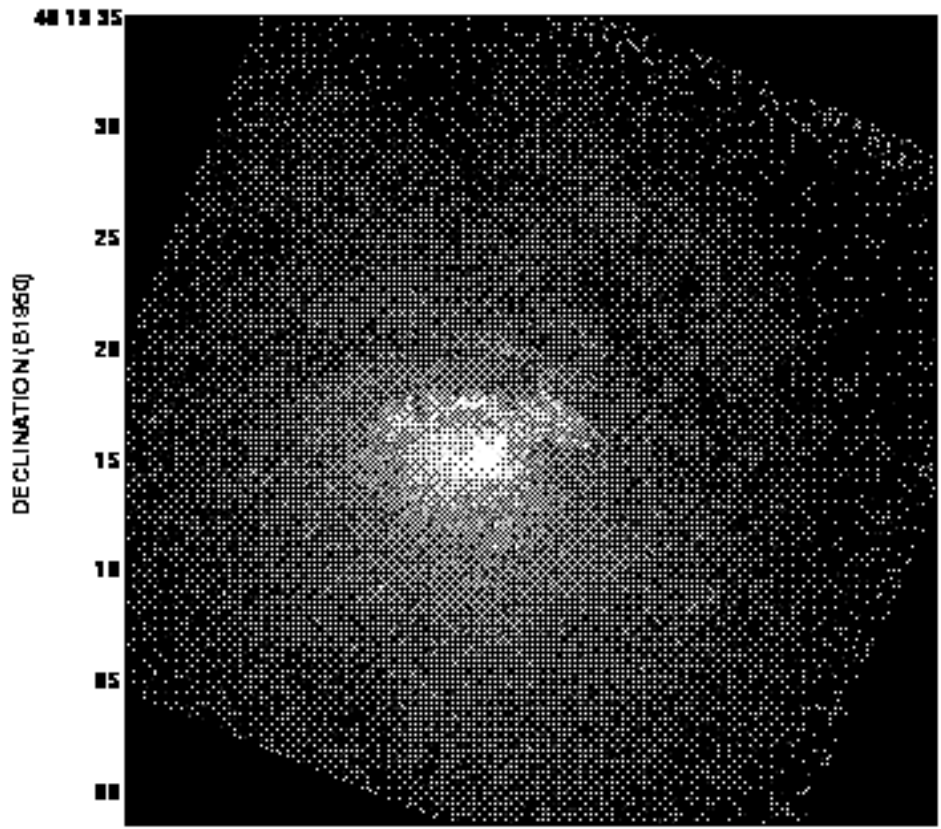
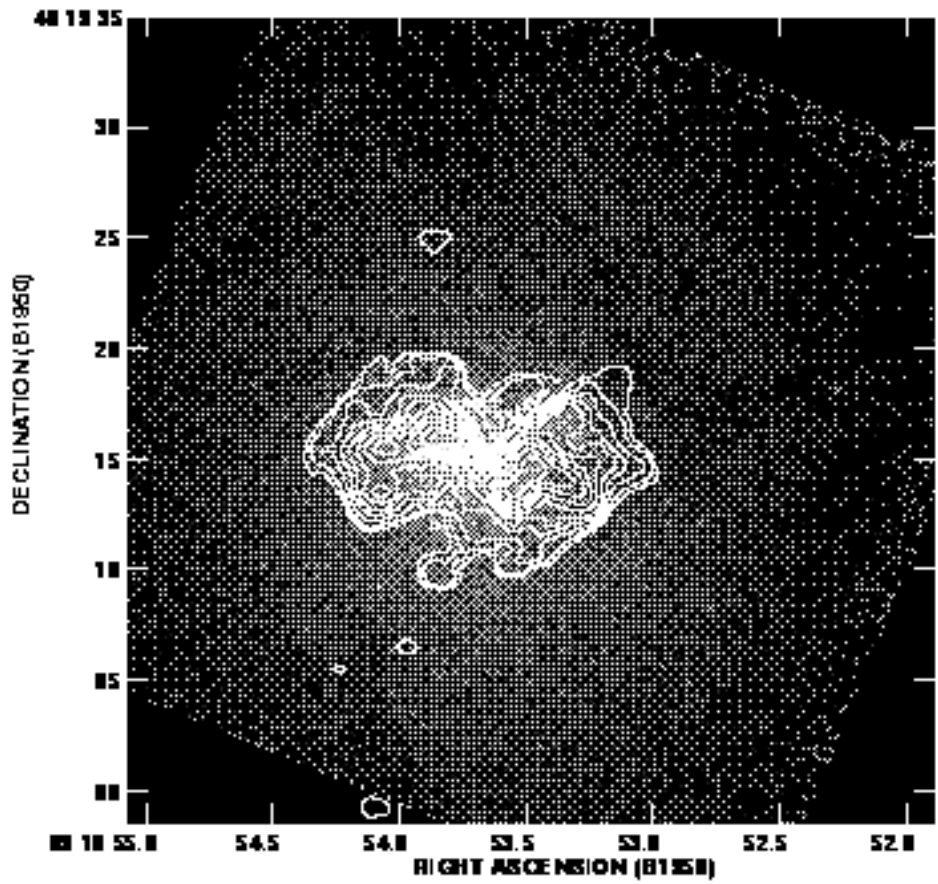
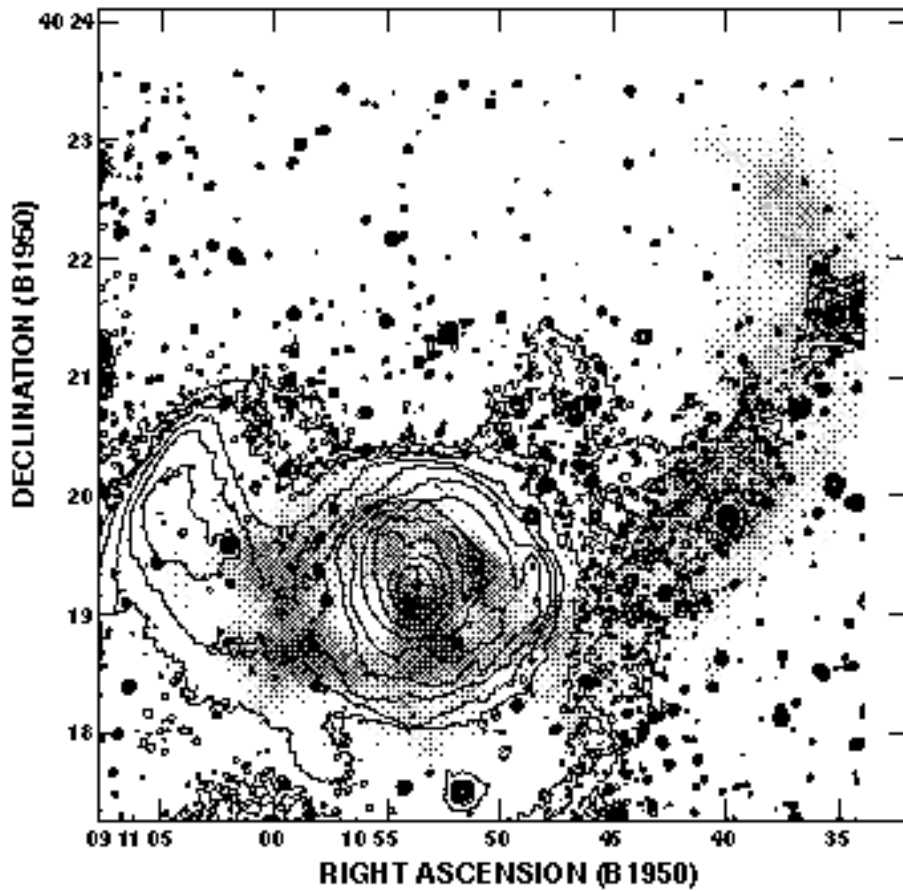
\end{document}